\documentclass[pra,twocolumn,superscriptaddress,showpacs,amsmath,amstex,amssymb,citeautoscript]{revtex4-1}

\usepackage[T1]{fontenc}
\usepackage[utf8]{inputenc}
\usepackage{lipsum}
\usepackage{amsmath}
\usepackage{amssymb}
\usepackage{bbm}
\usepackage{braket}
\usepackage{xcolor}
\usepackage{pifont}
\usepackage[mathscr]{euscript}
\usepackage[shortlabels]{enumitem}
\usepackage[justification=justified]{subcaption}
\usepackage{graphicx}
\usepackage{lipsum}
\allowdisplaybreaks
\usepackage{float}
\usepackage{graphicx}
\usepackage{dsfont}
\usepackage{comment}
\usepackage[colorlinks=true]{hyperref}  
\hypersetup{
    bookmarks=true,         
    unicode=false,          
    pdftoolbar=true,        
    pdfmenubar=true,        
    pdffitwindow=false,     
    pdfstartview={FitH},    
    pdftitle={Vestigial pairing},    
    pdfauthor={Poduval and Scheurer},     
    pdfsubject={},   
    pdfcreator={},   
    pdfproducer={}, 
    pdfkeywords={} {} {}, 
    pdfnewwindow=true,      
    colorlinks=true,       
    linkcolor=blue, 
    citecolor=blue,        
    filecolor=magenta,      
    urlcolor=blue           
} 
\newcommand{\bN}{\vec{N}}
\newcommand{\bd}{\vec{d}}
\newcommand{\bs}{\mathbf{s}}
\newcommand{\bS}{\mathbf{S}}
\newcommand{\bx}{\vec{x}}
\newcommand{\bB}{\vec{B}}
\newcommand{\bA}{\vec{A}}
\newcommand{\ba}{\vec{a}}
\newcommand{\bb}{\vec{b}}
\newcommand{\bD}{\mathbf{D}}

\newcommand{\bk}{\vec{k}}
\newcommand{\bq}{\vec{q}}

\newcommand{\equref}[1]{Eq.~(\ref{#1})}
\newcommand{\equsref}[2]{Eqs.~(\ref{#1}) and (\ref{#2})}
\newcommand{\secref}[1]{Sec.~\ref{#1}}
\newcommand{\figref}[1]{Fig.~\ref{#1}}

\newcommand{\tableref}[1]{Table~\ref{#1}}

\newcommand{\pdagger}{{\phantom{\dagger}}}
\newcommand{\diff}{\mathrm{d}}

\renewcommand{\approx}{\simeq}

\renewcommand{\vec}[1]{\boldsymbol{#1}}

\definecolor{wrongultramarine}{rgb}{1,0.5,0}

\begin{document}

\title{Vestigial singlet pairing in a fluctuating magnetic triplet superconductor: \\ Applications to graphene moir\'e systems}

\author{Prathyush P. Poduval}
\affiliation{Condensed Matter Theory Center, Department of Physics, University of Maryland, College Park, MD 20742, USA}

\author{Mathias S.~Scheurer}
\affiliation{Institute for Theoretical Physics, University of Innsbruck, Innsbruck A-6020, Austria}

\begin{abstract}
Motivated by the phenomenology of graphene moir\'e superlattices, we study a 2D model with strong tendencies towards both magnetism and triplet superconductivity. Individually, their respective order parameters, $\vec{N}$ and $\vec{d}$, cannot order at finite temperature. Nonetheless, the model exhibits a variety of vestigial phases, including charge-$4e$ superconductivity and broken time-reversal symmetry. Our main focus is on a phase characterized by finite $\vec{d} \cdot \vec{N}$, which has the same symmetries as the BCS state, a Meissner effect, and metastable supercurrents, yet rather different spectral properties: most notably, the suppression of the electronic density of states at the Fermi can resemble that of either a fully gapped or nodal superconductor, depending on parameters. This could provide a possible explanation for recent tunneling experiments in graphene moir\'e systems.
\end{abstract}

\maketitle

Strongly correlated systems often exhibit complex phase diagrams with multiple phases, characterized by long-range or quasi-long-range order (QLRO) of different order parameters. Aside from phase competition as a possible origin, a rich set of phases might also be understood as different manifestations of an underlying primary order---a concept often referred to as ``intertwined orders’’ \cite{fradkinColloquiumTheoryIntertwined2015}. For instance, thermal or quantum fluctuations can disorder a primary order parameter, while higher-order composite order parameters can still survive. An example of such a ``vestigial phase’’ \cite{nieQuenchedDisorderVestigial2014,fernandesIntertwinedVestigialOrder2019}, is the charge-$4e$ superconducting state that emerges when a charge-$2e$ pair density wave order parameter, $\Delta_{\vec{Q}}$, itself vanishes, yet $\Delta_{\vec{Q}}\Delta_{-\vec{Q}}$ does not \cite{bergCharge4eSuperconductivityPairdensitywave2009}; this and other forms of charge-$4e$ superconductivity have attracted a lot of attention \cite{fernandesChargeSuperconductivityMulticomponent2021, jianChargeSuperconductivityNematic2021, zengPhasefluctuationInducedTimeReversal2021,songPhaseCoherencePairs2022,maccariPossibleTimereversalsymmetrybreakingFermionic2022,chungBerezinskiiKosterlitzThoulessTransitionTransport2022,jiangCharge4eSuperconductors2017,liChargeSuperconductorWavefunction2022,gnezdilovSolvableModelCharge2022,curtisStabilizingFluctuatingSpintriplet2022,garaudEffectiveModelMagnetic2022,panFrustratedSuperconductivityCharge6e2022,yuNonuniformVestigialCharge4e2022,zhouChernFermiPocket2022}, in particular, as a result of recent experiments \cite{grinenkoStateSpontaneouslyBroken2021,geDiscoveryCharge4eCharge6e2022}.

\begin{figure}[b]
   \centering
    \includegraphics[width=\linewidth]{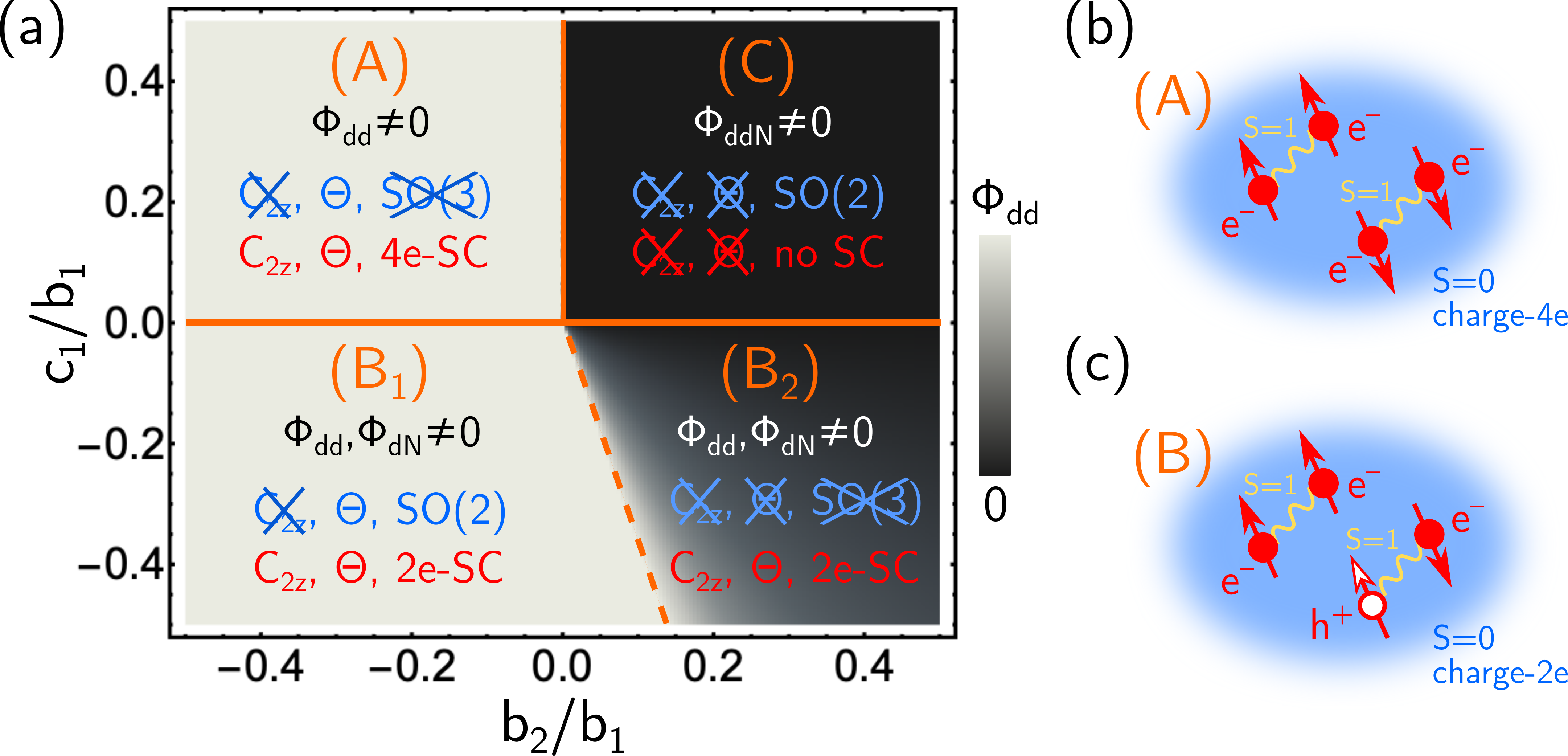}
    \caption{(a) Mean-field phase diagram for $r_d=r_N$, $b_3=b_1$, $c_2=0$, where we indicate the symmetries at $T=0$ (blue), those of the resulting vestigial phases at $T>0$ (red), and which composite order parameters are finite. Solid (dashed) orange lines are phase transitions at $T=0$ and $T>0$ (become a crossover at $T>0$). (b,c) illustrate the finite-$T$ pairing in phases (A) and (B) schematically.}
    \label{fig:MeanField}
\end{figure}

Another exciting recent development is the emergence of twisted graphene moir\'e superlattices as versatile playgrounds for strongly correlated physics \cite{andreiGrapheneBilayersTwist2020, balentsSuperconductivityStrongCorrelations2020}. These systems display a variety of different phases such as nematic \cite{kerelskyMaximizedElectronInteractions2019,caoNematicityCompetingOrders2021, rubio-verduMoireNematicPhase2022a} and density-wave order \cite{heChiralitydependentTopologicalStates2021, polshynTopologicalChargeDensity2022, siriviboonNewFlavorCorrelation2022}, different forms of magnetism \cite{sharpeEmergentFerromagnetismThreequarters2019,polshynElectricalSwitchingMagnetic2020,chenElectricallyTunableCorrelated2021, kuiriSpontaneousTimereversalSymmetry2022,linZerofieldSuperconductingDiode2022}, and, possibly unconventional \cite{kimEvidenceUnconventionalSuperconductivity2022, ohEvidenceUnconventionalSuperconductivity2021}, superconductivity \cite{caoUnconventionalSuperconductivityMagicangle2018}; magnetism and superconductivity appear in the same density range  \cite{wongCascadeElectronicTransitions2020, zondinerCascadePhaseTransitions2020,parkTunableStronglyCoupled2021,haoElectricFieldTunable2021,kimEvidenceUnconventionalSuperconductivity2022,ohEvidenceUnconventionalSuperconductivity2021,morissetteElectronSpinResonance2022} and recent experiments \cite{linZerofieldSuperconductingDiode2022,scammellTheoryZerofieldSuperconducting2022} demonstrate that they can coexist microscopically. Motivated by these observations, we here study the case of two primary order parameters: a fully gapped spin-triplet superconductor ($\vec{d}$) and, in line with the conclusions of \cite{lakePairingSymmetryTwisted2022, morissetteElectronSpinResonance2022},  magnetic order ($\vec{N}$) with antiparallel spins in the two valleys. 
At finite temperature, $T>0$, it must hold $\braket{\vec{d}}=\braket{\vec{N}}=0$, in two-dimensions (2D). However, there are several different vestigial phases, see \figref{fig:MeanField}(a), characterized by the composite order parameters $\phi_{dd}=\vec{d}\cdot\vec{d}$, $\phi_{dN} =\vec{d}\cdot\vec{N}$, and $\phi_{ddN} = i (\vec{d}^\dagger\times \vec{d})\cdot\vec{N}$. These include not only a charge-$4e$ superconductor \cite{xuTopologicalSuperconductivityTwisted2018,scheurerPairingGraphenebasedMoir2020}, see \figref{fig:MeanField}(b), but also a charge-$2e$ state, which has the same symmetries as and is, hence, adiabatically connected to the BCS state. However, it should primarily be thought as a condensate of three electrons and a hole, see \figref{fig:MeanField}(c), or, more formally, QLRO of $\phi_{dN}$. We develop a theory for this state and study its spectral properties at finite $T$, which are rather different from those of the BCS state. Depending on $T$ and $\phi_{dN}$, we obtain a low-energy suppression of the density of states (DOS) similar to a fully gaped or nodal state. This could provide an alternative explanation \cite{sukhachovAndreevReflectionScanning2022,lakePairingSymmetryTwisted2022,islamUnconventionalSuperconductivityPreformed2022a}  to the tunneling data of \cite{kimEvidenceUnconventionalSuperconductivity2022, ohEvidenceUnconventionalSuperconductivity2021}, which does not require any momentum dependence in the superconducting order parameter.

\vspace{1em}
\textit{Model.---}We consider a 2D model exhibiting both triplet superconductivity and magnetism, with three-component order parameter fields $\vec{d}$ (complex) and $\vec{N}$ (real), respectively. Denoting the electronic field operators of spin $s=\uparrow,\downarrow$ (Pauli matrices $\vec{s}$) and in valley $\tau=\pm$ (Pauli matrices $\vec{\tau}$) by $c_{k,s,\tau}$, where $k=(i\omega_n,\vec{k})$ comprises Matsubara frequencies and 2D momentum, they couple as
\begin{equation*}
    \mathcal{S}_c = \lambda \int_{k,q} \hspace{-0.2em} \left[ c^\dagger_{k-q} \vec{s}\vec{N}_q \tau_z c^\pdagger_{k} + ( c^\dagger_{k-q} \vec{s}\vec{d}_q is_y\tau_y c^\dagger_{-k}  +  \text{H.c.})\right].
\end{equation*}
Note that $\vec{N}$ couples anti-ferromagnetically in the two valleys; while  the ferromagnetic case---with $\tau_0$ instead of $\tau_z$ in the first term of $\mathcal{S}_c$ above---can be studied similarly, we focus on antiferromagnetism not only for concreteness here but also because recent microwave experiments \cite{morissetteElectronSpinResonance2022} and a systematic analysis \cite{lakePairingSymmetryTwisted2022} of multiple other experiments on graphene moir\'e systems favor this scenario.
The bare dynamics of $\vec{d}$ and $\vec{N}$ is governed by
\begin{equation*}
    \mathcal{S}_\chi = \int_q \left[ \chi_N^{-1}(q)\, \vec{N}_q\vec{N}_{-q} + \chi_{d}^{-1}(q)\, \vec{d}^\dagger_q\vec{d}_{q} \right].
\end{equation*} 
We take the susceptibilitites to be $\chi_\mu(q) =  \chi^0_\mu/(r_\mu +\Omega_n^2+ v_\mu^2 \vec{q}^2)$, $\mu=N,d$, where $q=(i \Omega_n,\vec{q})$ and $\Omega_n$ are bosonic Matsubara frequencies. The nature of the phase realized in the system depends crucially on the interactions between the bosonic fields. Up to quartic order, the local terms allowed by the symmetries listed in \tableref{SymmetriesAndReps} can be written as $\mathcal{S}_V=\int_x V(\vec{d}(x),\vec{N}(x))$ with 
\begin{equation*}
    V = b_1 (\vec{d}^\dagger\vec{d})^2 + b_2 |\vec{d}\vec{d}|^2 + b_3 \vec{N}^4 + c_1 |\vec{d}  \vec{N}|^2 + c_2 (\vec{d}^\dagger\vec{d})\vec{N}^2.
\end{equation*}
Finally, the bare electronic action is given by
\begin{equation*}
    \mathcal{S}_e = \int_k c^\dagger_{k,\tau,s} \left( -i\omega_n + \epsilon_{\tau\cdot \vec{k}} \right) c_{k,\tau,s}^\pdagger ,
\end{equation*}
where we already used that the band structures in the two valleys are related by time-reversal $\Theta$. 

\vspace{1em}
\textit{Mean-field and possible phases.---}To probe the possible phases, we start with a mean-field analysis with respect to $\vec{d}$ and $\vec{N}$. Absorbing the impact of the coupling to the electrons \footnote{See Appendix.} by a redefinition of the parameters of $V$, we obtain the four distinct zero-temperature phases labeled $(A)$, $(B_{1,2})$, and $(C)$ in \figref{fig:MeanField}(a), where we assumed that both $\braket{\vec{d}}$ and $\braket{\vec{N}}$ are non-zero and homogeneous. Using $\hat{\vec{e}}_{1,2,3} \in \mathbb{R}^3$ to denote orthogonal unit vectors, we have $\vec{N} = N_0 \hat{\vec{e}}_1$ and $\vec{d} = d_0 e^{i\alpha} \hat{\vec{e}}_2$ in phase (A), which breaks SO(3) completely, while $\Theta$ is preserved (in any gauge-invariant observable); as for any phase with $\braket{\vec{N}} \neq 0$, $C_{2z}$ is broken. In phase $(B_1)$, $\vec{N}$ and $\vec{d}$ are aligned; we, thus, obtain a residual spin-rotation symmetry SO(2) along that direction and $\Theta$ is preserved too. Beyond a critical value of $b_2$, an additional component with relative phase $\pi/2$ emerges in $\vec{d}$, defining phase $(B_2)$ where $\vec{N} = N_0 \hat{\vec{e}}_1$ and $\vec{d} = d_0 e^{i\alpha} (\hat{\vec{e}}_1 + i \eta \hat{\vec{e}}_2)$, with $0<\eta < 1$; this is a distinct phase as $\eta\neq 0$ breaks both the residual SO(2) spin symmetry and $\Theta$. Finally, phase (C) is characterized by $\vec{N} = N_0 \hat{\vec{e}}_1$ and $\vec{d} = d_0 e^{i\alpha} (\hat{\vec{e}}_2 + i \hat{\vec{e}}_3)$. Consequently, $\Theta$ is also broken but a residual SO(2) spin-symmetry remains.

\begin{table}[tb]
\begin{center}
\caption{Relevant symmetries $g$ and their action on the field operators. Here $R_{\vec{\varphi}}$ is the orthogonal matrix obeying $e^{-i\vec{\varphi}\cdot\vec{s}} \vec{s} e^{i\vec{\varphi}\cdot\vec{s}} = R(\vec{\varphi}) \vec{s}$. All symmetries are linear except for $\Theta$ which is anti-linear.}
\label{SymmetriesAndReps}\begin{ruledtabular}\begin{tabular}{ccccccc} 
$g$ & $c_{\vec{k}}$ & $\vec{N}$ & $\vec{d}$ & $\phi_{dd}$ & $\phi_{dN}$ & $\phi_{ddN}$ \\ \hline
$U(1)$ & $e^{i\varphi}c_{\vec{k}}$ & $\vec{N}$ & $e^{-2i\varphi}\vec{d}$ & $e^{-4i\varphi}\phi_{dd}$ & $e^{-2i\varphi}\phi_{dN}$ & $\phi_{ddN}$ \\
SO(3) & $e^{i\vec{\varphi}\cdot\vec{s}}c_{\vec{k}}$  & $R_{\vec{\varphi}} \vec{N}$ & $R_{\vec{\varphi}} \vec{d}$ & $\phi_{dd}$ & $\phi_{dN}$ & $\phi_{ddN}$ \\
$C_{2z}$ & $\tau_x c_{-\vec{k}}$  & $-\vec{N}$ & $-\vec{d}$ & $\phi_{dd}$ & $\phi_{dN}$ & $-\phi_{ddN}$  \\
$\Theta$ & $i s_y \tau_x c_{-\vec{k}}$  & $\vec{N}$ & $-\vec{d}^*$ & $\phi^*_{dd}$ & $-\phi^*_{dN}$ & $-\phi_{ddN}$
 \end{tabular}
\end{ruledtabular}
\end{center}
\end{table}

Importantly, $\braket{\vec{d}},\braket{\vec{N}} \neq 0$ is only possible and, thus, our discussion of symmetries is only valid for $T=0$ in 2D. To analyze the resulting vestigial phases at finite $T$, where SO(3) spin-rotation symmetry is preserved and $\braket{\vec{d}}=\braket{\vec{N}} = 0$, it is convenient to define the following composite order parameters $\phi_{dd} = \vec{d}\cdot \vec{d}$, $\phi_{dN} = \vec{d}\cdot \vec{N}$, and $\phi_{ddN} = i (\vec{d}^\dagger\times \vec{d}) \cdot \vec{N}$, with symmetry properties listed in \tableref{SymmetriesAndReps}. Crucially, all of them transform trivially under SO(3) spin-rotations and, hence, can exhibit long-range (in case of the last one) or QLRO (in case of the former two) at finite $T$. We indicate this in \figref{fig:MeanField}(a) for the different phases. This immediately tells us that, in spite of $\braket{\vec{d}}=0$, phase (A) transitions for finite $T$ into state where $\phi_{dd}$ has QLRO and, thus, constitutes a charge-$4e$ superconductor (as $\phi_{dN}=0$), which does not break $C_{2z}$ or $\Theta$ (as $\phi_{ddN}=0$); intuitively, one can think of this state as a condensate of four electrons forming a spin-singlet out of two triplets, see \figref{fig:MeanField}(b). At finite $T$, $(B_1)$ and $(B_2)$ will both preserve all normal-state symmetries 
and become the same phase, which we denote by $(B)$ in the following. 
It is characterized by QLRO not only in $\phi_{dd}$ but also in $\phi_{dN}$; as the latter has charge $2e$, it is a charge-$2e$ superconductor and adiabatically connected to the conventional BCS state. Nonetheless, in our current description, this state should rather be thought of as the condensation of three electrons and a hole, see \figref{fig:MeanField}(c), consisting of a pair of electrons in a triplet state forming a singlet with a spin-$1$ particle-hole excitation. In fact, we will see below that it exhibits spectral properties rather different from those of the BCS state at finite $T$.
Finally, while phase $(C)$ does not exhibit any vestigial pairing at $T>0$, it will have long-range order in $\phi_{ddN}$ and, as such, continues to break both $C_{2z}$ and $\Theta$.

\vspace{1em}
\textit{Theory for phase (B).---}As $c_1 < 0$ is found when the coefficients in $V$ are computed by integrating out electrons \cite{Note1}, we next focus on phase (B). To obtain an efficient description of this phase that properly captures the preserved SO(3) symmetry at finite temperature, we first decouple the four terms in $V$ using four Hubbard-Stratonovich fields, $\psi_d$ for $\vec{d}^\dagger \vec{d}$, $\psi_N$ for $\vec{N}^2$, $\phi_d$ for $\vec{d}\cdot\vec{d}$, and $\phi_{dN}$ for $\vec{d}\cdot\vec{N}$. We treat them on the saddle-point level, which becomes exact in the limit where the number of components of $\vec{d}$ and $\vec{N}$ is taken to be infinitely large \cite{fernandesPreemptiveNematicOrder2012}. The saddle point values of  $\psi_d$ and $\psi_N$ will in general be non-zero, which we absorb into a redefinition of $r_{d,N}$. Then, the effective action for phase (B) becomes $\mathcal{S}_{\text{eff}} = \mathcal{S}_\chi + \mathcal{S}_e + \mathcal{S}_c + \mathcal{S}_{\phi}$ where 
\begin{equation}
    \mathcal{S}_{\phi} = \int_{q} \left[\phi^0_{dN} \,\vec{d}_q\cdot \vec{N}_{-q} + \phi^0_{dd} \,\vec{d}_q\cdot \vec{d}_{-q} + \text{H.c.}  \right]. \label{SphiAction}
\end{equation}
While generically both saddle point values $\phi^0_{dN}$ and $\phi^0_{dd}$ are expected to be non-zero simultaneously in phase (B), we take $\phi^0_{dd} \rightarrow 0$ and $\phi_{dN}^0 \equiv  \phi_0 \neq 0$ for the following explicit calculations. Setting $\phi^0_{dd} = 0$ does not change any symmetries of the phase, allows for a more compact discussion of the results, and can formally be seen as the large $b_2$ limit of the theory where $\phi^0_{dd}$ is suppressed [cf.~\figref{fig:MeanField}(a)]. More generally than the derivation of $\mathcal{S}_{\text{eff}}$ via Hubbard-Stratonovich transformations, it can also be thought of as the simplest field theory capturing the key aspects of phase (B) in \figref{fig:MeanField}(a) at finite $T$.

\begin{figure}[tb]
   \centering
    \includegraphics[width=\linewidth]{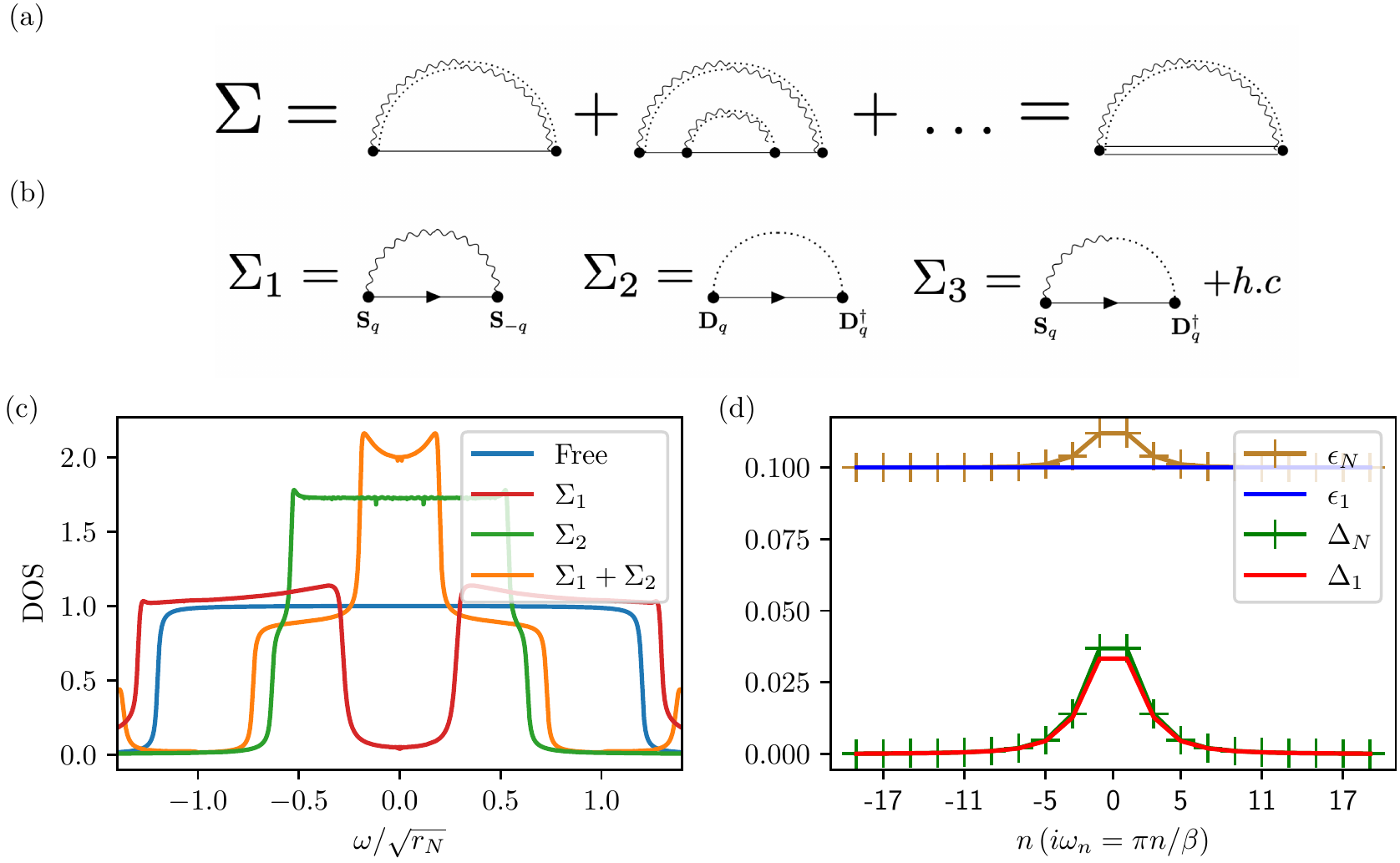}
    \caption{Diagrams contributing to the fermionic self energy $\Sigma$ (a) in the matrix-large-$N$ limit defined in the main text and (b) to first order. (c) Impact of spin $(\Sigma_1)$ and triplet fluctuations $(\Sigma_2)$ on the constant DOS (blue) of a 2D band with finite bandwidth. (d) Comparing the first order solution  $(\epsilon_1,\Delta_1)$ and self consistent solution $(\epsilon_N,\Delta_N)$ for $G=i\omega-\epsilon(i\omega)\gamma_z+\Delta(i\omega)\gamma_y$ for $\mathcal{S}_2$ only (both without momentum integration). We use $\epsilon/\sqrt{r_N}=0.1,\phi_0/r_N=0.5$.}
    \label{fig:DOSContr}
\end{figure} 

\vspace{1em}
\textit{Electronic self energy.---}To compute the spectral properties of the electrons within this model, we employ a large-$N$ technique similar to \cite{fitzpatrickNonFermiliquidBehaviorLarge2014, wermanMottIoffeRegelLimitResistivity2016}: we add extra indices to the electrons and bosons, $c_{k,\tau,s}\rightarrow c_{k,\tau,s,a}$, $\vec{d}_{ab} \rightarrow \vec{d}_{ab}$ and similarly for $\vec{N}$, where $a,b=1,2,\dots, N$, which are contracted in all terms of $\mathcal{S}_{\text{eff}}$ so as to ensure $O(N)$ symmetry. In the limit $N\rightarrow \infty$, the electronic self-energy $\Sigma$ is given by the ``rainbow diagrams'' \cite{fitzpatrickNonFermiliquidBehaviorLarge2014, wermanMottIoffeRegelLimitResistivity2016} shown in \figref{fig:DOSContr}(a). In our case, however, $\Sigma$ involves both normal and anomalous contributions as a result of the anomalous bosonic term $\propto \phi_0$ in \equref{SphiAction}. To make this more explicit, we integrate out the bosons, yielding the effective fermionic interactions $\mathcal{S}_{\text{int}}=\mathcal{S}_1+\mathcal{S}_2$ with
\begin{subequations}\begin{align}
    \mathcal{S}_1&=-\int_q \frac{\lambda^2}{M_q}\left(\frac{\chi_{d}^{-1}}{4}\bS_q\cdot\bS_{-q}  +\chi_N^{-1} \bD_q \cdot \bD_q^\dagger\right),\\
    \mathcal{S}_2&=-\frac12\int_q \frac{\lambda^2}{M_q}\left(\phi_0\,\bS_q\cdot \bD_q^\dagger+ \phi_0^* \, \bD_q\cdot\bS_{-q}\right), \label{AnomalousInteractionTerm}
\end{align}\end{subequations}
where $M_q=\chi_d^{-1}\chi_N^{-1}-|\phi_0|^2$ and $\bS_q=\int_k c^\dagger_{k+q} \bs \tau_zc_k$, $\bD_q=\int_k c^\dagger_{k+q} \bs is_y\tau_yc^\dagger_{-k}$. The two terms in $\mathcal{S}_1$ describe spin and superconducting triplet fluctuations, respectively; their associated self-energy contributions are normal in the sense that $U(1)$ symmetry is preserved, with leading terms represented by the first two diagrams $\Sigma_{1,2}$ in \figref{fig:DOSContr}(b). Conversely, $\mathcal{S}_2$ breaks $U(1)$ symmetry, when $\phi_0$ attains a mean-field value, and results in an anomalous contribution to the self-energy, with leading term given by the last diagram $\Sigma_3$ in \figref{fig:DOSContr}(b).

To represent the diagrams algebraically, we shift to the Bogoliubov-de Gennes basis $(c_{q,+}, is_yc_{-q,-}^\dagger)^T$, with Pauli matrices $\gamma_i$ acting on this space. In this basis, the free Green's function is $G_0(i\omega,\epsilon) = i\omega-\epsilon\gamma_z$. Up to first order in $\lambda^2$, the spin-spin self energy term can be written as $\Sigma_1(k)=3\lambda^2\int_q  \frac{\chi_{d}^{-1}(q)}{2M_q} G_0(i\omega+i\Omega,\epsilon_{\bk+\bq})$, while the triplet-triplet term is $\Sigma_2(k)=12 \lambda^2\int_q  \frac{\chi_{N}^{-1}(q)}{M_q} G_0(i\omega+i\Omega,-\epsilon_{\bk+\bq})$. After performing a gauge transformation to make $\phi_0$ real, the anomalous term from the spin-triplet interaction is given by
\begin{align}
    \Sigma_3(k)=3\phi_0\int_q\frac{\lambda^2}{M_q} \{\gamma_y,\gamma_z G_0(i\omega+i\Omega,\epsilon_{\bk+\bq})\}. \label{SelfEnergy3}
\end{align}
For concreteness and since spin fluctuations are believed to occur already at higher energies than superconducting fluctuations in graphene moir\'e systems \cite{wongCascadeElectronicTransitions2020, zondinerCascadePhaseTransitions2020}, we focus on $r_d > r_N$; we will use  $r_d/r_N=9,v_d^2/v_N^2=8$, $\chi_N^0=\chi_d^0$, and set $\chi_\mu^0=1$ by rescaling of the fields.

\vspace{1em}
\textit{Density of states.---}Figure \ref{fig:DOSContr}(c) shows the effect of the normal contributions of the self energy $\Sigma_{1,2}$ on the DOS of a 2D parabolic band. The effect of $\Sigma_1$ is to push the peak of the free spectral function at energy $\epsilon$ away from $\omega=0$. This results in the opening of a gap (which can be soft depending on the parameter regime), very similar to fluctuating anti-ferromagnetism discussed in the cuprates \cite{kyungPseudogapSpinFluctuations2004, vilkNonPerturbativeManyBodyApproach1997,scheurerTopologicalOrderPseudogap2018}. $\Sigma_2$ on the other hand has the opposite effect, where it pushes states \textit{towards} $\omega=0$. This is because $\Sigma_1$ and $\Sigma_2$ have the exact same functional form with one key difference: $\epsilon_{\bk+\bq}$ of $\Sigma_1$ is replaced by $-\epsilon_{\bk+\bq}$ in $\Sigma_2$. 
The effect of the total normal self energy $\Sigma_1+\Sigma_2$ is to enhance the DOS in the vicinity of the Fermi level, see \figref{fig:DOSContr}(c).
The anomalous contribution $\Sigma_3$ does not interfere in these effects since it occurs in the $\gamma_y$ channel. 
The role of $\Sigma_1+\Sigma_2$ can, thus, be intuitively thought of as providing a renormalized DOS in the normal state on top of which the anomalous $\Sigma_3$ opens up a gap.
We have checked \cite{Note1} by numerically solving the self-consistency equation for the self-energy [\figref{fig:DOSContr}(a)] in the limit (of large $v_\mu$) where only the $\bq=0$ term of the momentum sum contributes that higher-order corrections do not change our results qualitatively for small $\phi_0$. For instance, \figref{fig:DOSContr}(d) shows the numerical solution for the Green's function $G=i\omega-\epsilon(i\omega)\gamma_z+\Delta(i\omega)\gamma_y$ in Matsubara space upon including the effect of $\mathcal{S}_2$; the difference to the first-order result is small.

\begin{figure}[tb]
   \centering
    \includegraphics[width=\linewidth]{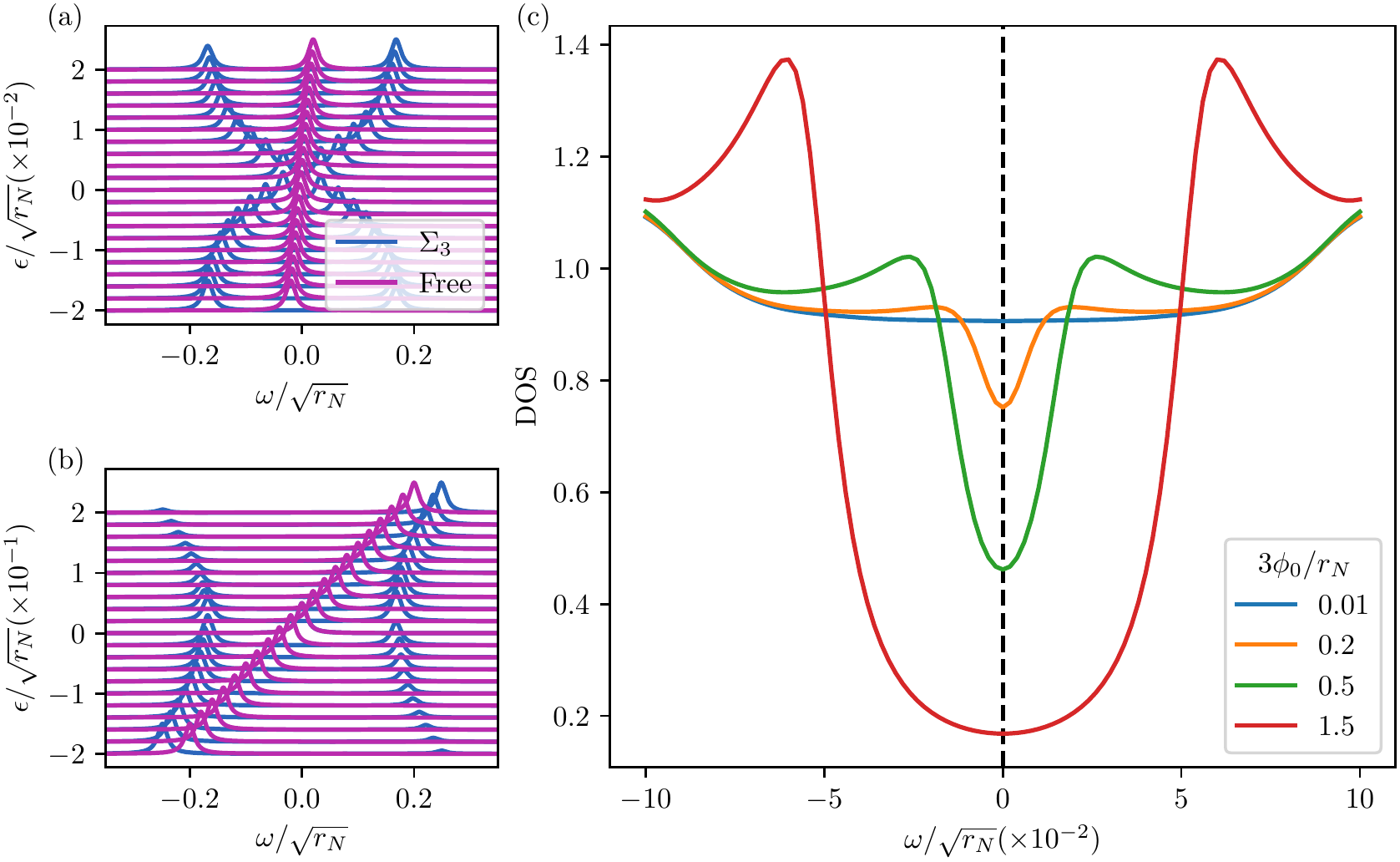}
    \caption{Spectral weight as a function of $\omega$ with (blue) and without (purple) $\Sigma_3$ (a) close to $\epsilon_{\vec{k}}=0 $ and (b) including a larger energy range; in both cases, we focus on the $\vec{q}=0$ contribution (see text). (c) The effect of all three self energy contributions $ \Sigma_1+\Sigma_2+\Sigma_3$ (including the momentum integration) on the DOS. For small $\phi_0$, there is suppression of the DOS at $\omega=0$ which resembles the V-shaped DOS of a nodal state. For large $\phi_0$, the gap resembles a hard BCS gap.}
    \label{fig:DOSFull}
\end{figure} 

To gain intuition for the impact of $\Sigma_3$ on the DOS, we first focus again on the $\bq=0$ term of the momentum sum in \equref{SelfEnergy3}. In this limit, one can easily see \cite{Note1} that $\Sigma_3$ vanishes linearly in $\epsilon_{\vec{k}}$ for small energies. Since $\Sigma_3$ is in the $\gamma_y$ channel, the effect of any non-zero value is to generically open a gap. As a result of the linear behavior, the states exactly at zero energy are unaffected, but slightly away from it, the states get pushed away to higher energy; this is clearly visible in \figref{fig:DOSFull}(a). 
In contrast, for large energies, $\Sigma_3$ is readily seen to tend to zero. The spectral function, thus, remains asymptotically unaffected, as can be seen in \figref{fig:DOSFull}(b). Taken together, we expect the DOS to be reduced (but not fully suppressed for small $\phi_0$) in an energy range around the Fermi level, exhibiting an enhancement with respect to its normal-state value at intermediate energies, and then approaching the normal-state limit at larger energies.

\begin{figure}[tb]
   \centering
    \includegraphics[width=\linewidth]{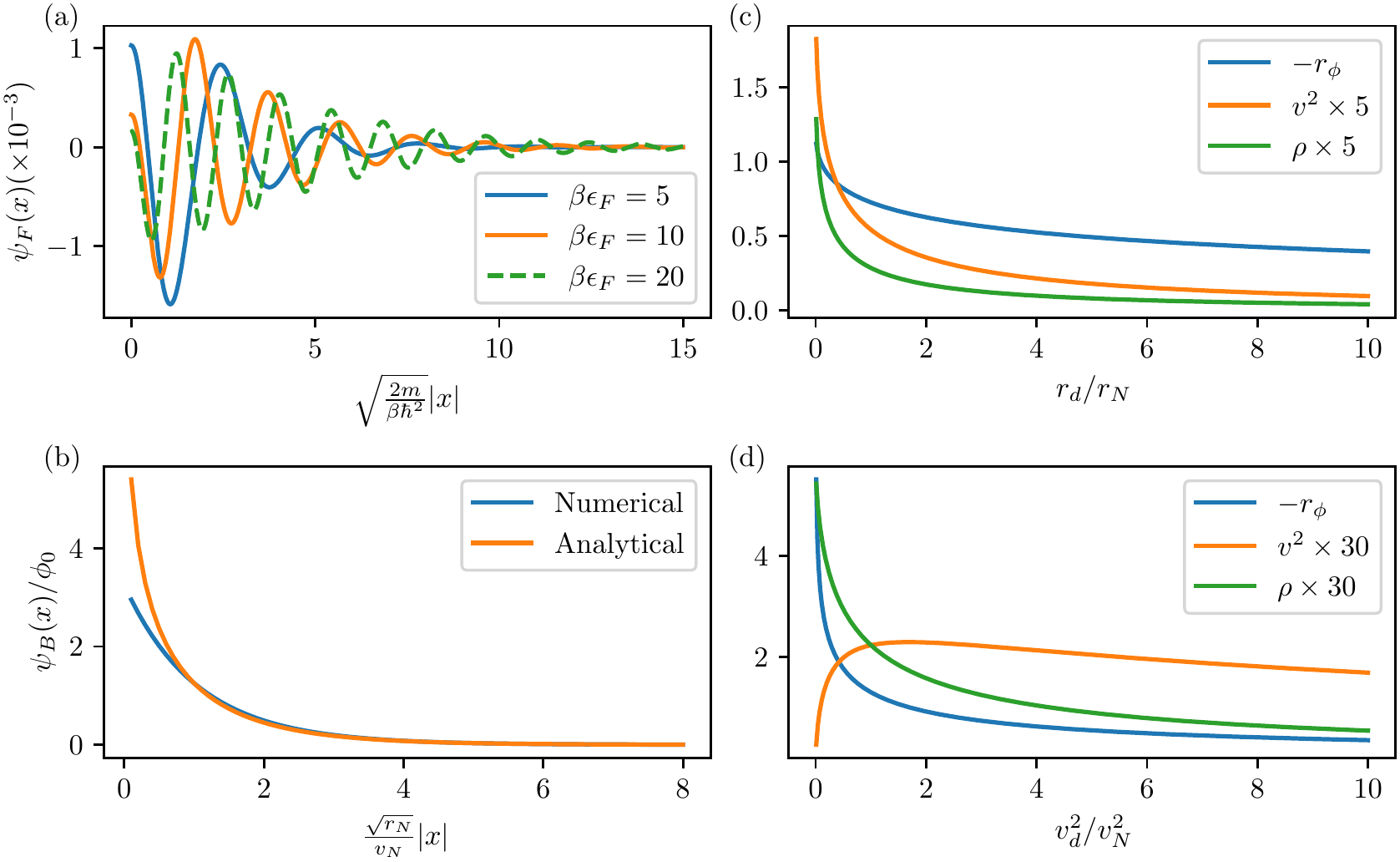}
    \caption{(a) The fermionic and (b) the bosonic ODLRO ``macroscopic wavefunction''. The mass $r_{\phi}$ [in units of $r_N^{-1/2}v_N^{-2}$], superfluid density $\rho$ [$r_N^{-3/2}v_N^{-2}$], and velocity $v^2$ [$r_N^{-3/2}$] of $\mathcal{S}_{\text{GL}}$ in \equref{SphiGL} as a function of $r_d$ and $v_d^2$ are shown in (c) and (d), respectively.}
    \label{fig:ODLRO}
\end{figure} 

To demonstrate this explicitly beyond the simple $\vec{q}=0$ limit, we approximate $\epsilon_{\vec{k}+\vec{q}} \approx \epsilon_{\vec{k}}+v_F q_{\parallel}+\vec{q}^2/(2m)$, where $q_{\parallel}$ is the component of $\vec{q}$ along $\vec{k}$, and numerically evaluate the momentum integrals to find the total self energy $\Sigma=\Sigma_1+\Sigma_2+\Sigma_3$.
Choosing $v_F=1.5{v_N}, 2m=\sqrt{r_N}/v_N^2$ for concreteness, \figref{fig:DOSFull}(c) shows the resulting DOS. As expected, we see that there is a suppression of the DOS. However, for small values of $\phi_0$, the resulting DOS has a V-shaped behavior, which is typically only seen in nodal states (with either nodal lines or points). Recall that the superconducting phase in our model is symmetry-equivalent to a conventional BCS state and that the triplet superconductor that arises at $T=0$ in phase (B) will be fully gapped. For larger $\phi_0$, the gap at $\omega=0$ increases, and resembles a hard BCS gap. 
The suppression of the DOS $\rho_F$ at $\omega=0$ can be estimated analytically at finite temperature by again taking the limit (of large $v_\mu$) where the integration over $\vec{q}$ can be replaced by an evaluation at $\vec{q}=0$; we find
\begin{equation}
    \frac{\rho_F(\phi_0)}{\rho_F(\phi_0=0)} = \frac{1}{\sqrt{1+\alpha^2}}, \quad \alpha=\frac{3\phi_0 \lambda^2r_N}{2Tv_N^2(r_dr_N-\phi_0^2)}. \label{SuppressionAndAlpha}
\end{equation}
Note that $\phi_0^2$ is bounded above by $r_dr_N$, at which point the bosonic fields would condense and continuous symmetries would be broken, which cannot happen at finite $T$. As $\phi_0$ increases, $\alpha$ increases the suppression of the DOS, and near the instability point of $\phi_0^2=r_dr_N$, there are no states near the Fermi surface. 

To complement this analysis, we have also studied the Hamiltonian associated with setting $q=0$ in \equref{AnomalousInteractionTerm} within self-consistent Hartree-Fock, only allowing for spin-rotation invariant operators to condense \cite{Note1}. For small $\alpha$, one also finds only a partial suppression of the low-energy spectral weight, akin to \equref{SuppressionAndAlpha}; including higher-order corrections leads to a hard gap for $\alpha \geq 1$.

\vspace{1em}
\textit{Electromagnetic response.---}We will finally demonstrate that the superconducting phase characterized by $\phi_0 \neq 0$ has the same electromagnetic phenomenology as BCS superconductors, despite the unusual electronic spectral properties. To this end, we study off-diagonal long-range order (ODLRO) \cite{penroseBoseEinsteinCondensationLiquid1956, penroseCXXXVIQuantumMechanics1951,yangConceptOffDiagonalLongRange1962} which implies the Meissner effect \cite{sewellOffdiagonalLongrangeOrder1990}, flux quantization \cite{niehOffdiagonalLongrangeOrder1995}, Josephson effect and persistent currents \cite{sewellOffdiagonalLongRange1997}. First focusing on the electrons, we show that $\braket{c^\dagger_{s_1,+}(\vec{x}_1)c^\dagger_{s_2,-}(\vec{x}_2)c^\pdagger_{s_2',-}(\vec{x}_2')c^\pdagger_{s_1',+}(\vec{x}_1')} \rightarrow n_0 (\Psi^*_{\text{F}}(\vec{x}_{12}))_{s_1,s_2} (\Psi_{\text{F}}(\vec{x}'_{12}))_{s_1',s_2'}$, with $\Psi_{\text{F}} \neq 0$, as $|\vec{x}_j-\vec{x}'_j| \rightarrow \infty$ at finite $\vec{x}_{12}=\vec{x}_1-\vec{x}_2$ and $\vec{x}'_{12}=\vec{x}'_1-\vec{x}'_2$, to leading (first) order in $\phi_0$; as non-zero $\Psi_{\text{F}}$ to linear order in $\phi_0$ implies that it cannot vanish identically for generic $\phi_0$, this is sufficient to show the presence of ODLRO. We find the ``macroscopic wave function'' to be a singlet, $\Psi_{\text{F}}(\vec{x}) = is_y \psi_{\text{F}}(\vec{x})$, as expected since spin-rotation symmetry is preserved at finite $T$, with $\psi_{\text{F}}(\vec{x})$ shown in \figref{fig:ODLRO}(a).
Alternatively, one can demonstrate ODLRO to arbitrary order in $\phi_0$, by focusing on the bosons: to zeroth order in $\lambda$, we find $\braket{(\vec{d}^\dagger(\vec{x}_1)\vec{N}(\vec{x}_2))(\vec{d}(\vec{x}'_1)\vec{N}(\vec{x}'_2))} \rightarrow  \psi^*_{\text{B}}(\vec{x}_{12}) \psi_{\text{B}}(\vec{x}'_{12})$ as $|\vec{x}_j-\vec{x}'_j| \rightarrow \infty$, with $\psi_{\text{B}}(\vec{x})$ plotted in \figref{fig:ODLRO}(b) along with an analytic asymptotic form for large $\vec{x}$; in \cite{Note1}, we show that this leads to the same constraints as the conventional form of bosonic ODLRO \cite{penroseBoseEinsteinCondensationLiquid1956, penroseCXXXVIQuantumMechanics1951}.
Finally, the connection to the textbook theory of superconductivity can be made more explicit by deriving the analogue of the time-dependent Ginzburg-Landay theory: we reinstate fluctuations via $\phi_0 \rightarrow \phi(\vec{x},\tau)$ and integrate out all other degrees of freedom yielding
\begin{equation}
    \mathcal{S}_{\text{GL}} = \int_{\vec{x},\tau} \hspace{-0.3em}\left[  \rho  |D_\tau \phi|^2 + (r_\phi+|c_1|^{-1})  | \phi|^2 + v^2\, |\vec{D}\phi|^2   \right] \label{SphiGL}
\end{equation}
to leading order in $\phi$ and gauge-covariant derivatives $(D_\tau,\vec{D})_\mu = \partial_\mu -i2e A_\mu$. For demonstration purposes, we evaluated the coefficients in $\mathcal{S}_{\text{GL}}$ to leading (zeroth) order in $\mathcal{S}_c$ and find $\rho,v_\phi >0$ and $r_\phi<0$ for low $T$ [see \figref{fig:ODLRO}(c,d)]; the state with QLRO in $\phi_0$ thus corresponds, as usual, to the Higgs phase, with Meissner effect and massive Higgs mode, but without Goldstone modes.

\vspace{1em}
\textit{Conclusion.---}We have studied the finite-$T$ vestigial phases, see \figref{fig:MeanField}(a), associated with two primary order parameters, $\vec{d}$ and $\vec{N}$, describing a fully gapped triplet superconductor and spin magnetism, respectively.  
A crucial result is the DOS of phase $(B_{1,2})$ in \figref{fig:DOSFull}(c): varying $\phi_0$ changes the low-energy DOS from partial suppression, akin to that of a nodal superconducting state, to a hard gap. As $\phi_0$ is expected to change with electron filling, this could explain the tunneling data in \cite{kimEvidenceUnconventionalSuperconductivity2022, ohEvidenceUnconventionalSuperconductivity2021}. 
We finally point out that the suppression of $\vec{N}$ would immediately also suppress $\phi_0$ in our model and could, therefore, explain why superconductivity is connected to the reset behavior in trilayer graphene \cite{parkTunableStronglyCoupled2021,haoElectricFieldTunable2021,kimEvidenceUnconventionalSuperconductivity2022,ohEvidenceUnconventionalSuperconductivity2021}.

\vspace{1em}
\textbf{Acknowledgments.} We thank Rafael Fernandes, Victor Gurarie, Peter Orth, and Subir Sachdev for fruitful discussions on the project and Jakob Wessling for a related collaboration. M.S.S.~acknowledges funding by the European Union (ERC-2021-STG, Project 101040651---SuperCorr). Views and opinions expressed are however those of the authors only and do not necessarily reflect those of the European Union or the European Research Council Executive Agency. Neither the European Union nor the granting authority can be held responsible for them. $P^3$ acknowledges support by the Laboratory for Physical Sciences through the Condensed Matter Theory Center.

\bibliography{VestigialSC}

\onecolumngrid

\begin{appendix}

\section{Mean-field form of the bosonic interactions}
In the main text, we view the field theory defined by the action $\mathcal{S} = \mathcal{S}_e +\mathcal{S}_\chi + \mathcal{S}_c + \mathcal{S}_V$ as an effective low-energy theory that arises when high-energy electronic degrees of freedom have already been integrated out. Due to the symmetry and locality constraints, it only depends on a few parameters, $r_\mu$, $v_\mu$, $b_{1,2,3}$, $c_{1,2}$. As can be seen in \figref{fig:MeanField}(a), in particular, (the sign of) the parameters $c_1$ and $b_2$ entering $V$ crucially determine the phase of the system. We here provide an estimate for these parameters using mean-field theory. To this end, we replace the bosonic fields by classical homogeneous and time-independent vectors, $\vec{N}_q \rightarrow \delta_{q,0}\vec{N}$,  $\vec{d}_q \rightarrow \delta_{q,0}\vec{d}$, in $\mathcal{S}_e +\mathcal{S}_\chi + \mathcal{S}_c$; this yields
\begin{equation}
    \mathcal{S}_{\text{HE}} = \int_k c^\dagger_{k,\tau,s} \left( -i\omega_n + \epsilon_{\tau\cdot \vec{k}} \right) c^\pdagger_{k,\tau,s} + \lambda \int_{k}\left[ c^\dagger_{k} \vec{s}\cdot \vec{N} \tau_z c^\pdagger_{k} + ( c^\dagger_{k} \vec{s}\cdot\vec{d} \, is_y\tau_y c^\dagger_{-k}  + \text{H.c.})\right] + \text{const.},
\end{equation}
which we now view as our full action, also containing the high-energy degrees of freedom. Integrating out the electronic degrees of freedom and expanding the resulting action in terms of $\vec{N}$ and $\vec{d}$ to quartic order, one obtains exactly the same terms as in $V$ defined in the main text, as expected by symmetry. Moreover, one finds
\begin{equation}
    c_1 = b_2 = -b_1/2 < 0, \quad \text{with} \quad b_1 = 32 \, \lambda^4 T\sum_{\omega_n}\int\frac{\diff^2 \vec{k}}{(2\pi)^2} \frac{1}{(\omega_n^2 +\epsilon_{\vec{k}}^2)^2} >0.
\end{equation}
As stated in the main text, this places us into phase (B). We note, however, that fluctuation corrections to mean field can modify the values of these coupling constants significantly \cite{fernandesNematicityProbeSuperconducting2013, koziiNematicSuperconductivityStabilized2019,scheurerPairingGraphenebasedMoir2020}. For instance, ferromagnetic fluctuations can change the sign of $b_2$ to positive values \cite{scheurerPairingGraphenebasedMoir2020}.

\section{Evaluation of the self-energies at leading order}
In this section, we show the evaluation of the self energies up to first order in perturbation theory. We first evaluate the anomalous part of the self energy, $\Sigma_3$ in \figref{fig:DOSContr}(b), which is contributed by the anomalous term of the action given by
\begin{align}
    \mathcal{S}_2&=-\frac12\int_q \frac{\lambda^2}{\chi_{d}^{-1}\chi_N^{-1}-|\phi_0|^2}\left(\phi_0\bS_q\cdot \bD_q^\dagger+ \phi_0^* \bD_q\cdot\bS_{-q}\right).
\end{align}
In the following, we work in the $\begin{pmatrix}c_{q,+}& is_yc_{-q,-}^\dagger\end{pmatrix}^T$ Bogoliubov-de Gennes basis, with the Pauli matrices $\gamma_i$ acting on it. The free Green's function then reads as $G_0^{-1}(k)=i\omega-\epsilon_{\bk}\gamma_z$. Choosing $\phi_0$ to be real, we have
\begin{align}
    \Sigma_3 = 3\int_q \frac{\phi_0\lambda^2}{M_q} \left(\gamma_yG_{0,k+q}\gamma_z+\gamma_zG_{0,k+q}\gamma_y\right)  = 6\int_q\frac{\phi_0\lambda^2}{M_q} \frac{\epsilon_{\bk+\bq}}{(i\omega+i\Omega)^2-\epsilon_{\bk+\bq}^2}\gamma_y,
\end{align}
where
\begin{align}
    M_q &= \chi_{N}^{-1}\chi_{d}^{-1}-\phi_0^2 =\left(-\left(i\Omega\right)^2+r_N+v_N^2\bq^2\right)\left(-\left(i\Omega\right)^2+r_d+v_{d}^2\bq^2\right)-\phi_0^2 \\
    &= ((i\Omega)^2-E_+^2)((i\Omega)^2-E_-^2),
\end{align}
with $E_{\pm}^2= \frac{g_d+g_N\pm\sqrt{(g_d-g_N)^2+4\phi_0^2}}{2}$, and $g_\mu=r_\mu+v_\mu^2\bq^2$. Thus,
\begin{align}
    \Sigma_3  = 6\phi_0\lambda^2\int_{\bq}T\sum_{i\Omega\in \text{Bosonic}}\frac{1}{\left(\left(i\Omega\right)^2-E_+(\bq)^2\right)\left(\left(i\Omega\right)^2-E_-(\bq)^2\right)} \frac{\epsilon_{\bk+\bq}}{(i\omega+i\Omega)^2-\epsilon_{\bk+\bq}^2}\gamma_y.
\end{align}

The Matsubara sum can be evaluated using 
\begin{align}
    f(i\omega,\epsilon)=&T\sum_{i\Omega}\frac{1}{((i\Omega)^2-E_+^2)((i\Omega)^2-E_-^2)}\frac{1}{i\omega+i\Omega-\epsilon}\\ &=\frac12\frac{1}{E_+^2-E_-^2}\left(\frac{1}{E_+}\left(K(i\omega,\epsilon,E_+)-K(i\omega,\epsilon,-E_+)\right)-\frac{1}{E_-}\left(K(i\omega,\epsilon,E_-)-K(i\omega,\epsilon,-E_-)\right)\right), \\
    K(i\omega,\epsilon,E) &= \frac{n_f(\epsilon)+n_B(-E)}{E+\epsilon-i\omega},\label{eq:keq}
\end{align}
where $n_{f/B}(\epsilon)=\frac{1}{e^{\beta\epsilon}\pm1}$.
Thus we get,
\begin{align}
    \Sigma_3(k)=3\phi_0\lambda^2\int_{\bq} \left(f(i\omega,\epsilon_{\bk+\bq})-f(i\omega,-\epsilon_{\bk+\bq})\right)\gamma_y,
\end{align}
where we performed a partial fraction decomposition of $\frac{2\epsilon_{\bk+\bq}}{(i\omega+i\Omega)^2-\epsilon_{\bk+\bq}^2}=\frac{1}{i\omega+i\Omega-\epsilon_{\bk+\bq}}-\frac{1}{i\omega+i\Omega+\epsilon_{\bk+\bq}}$ to arrive at the expression.

The normal part of the self energy, $\Sigma_{1,2}$ in \figref{fig:DOSContr}(b), is contributed by the following term of the action
\begin{align}
    \mathcal{S}_1&=-\int_q \frac{\lambda^2}{\chi_{d}^{-1}\chi_N^{-1}-|\phi_0|^2}\left(\frac{\chi_{d}^{-1}}{4}\bS_q\cdot\bS_{-q}  +\chi_N^{-1} \bD_q \cdot \bD_q^\dagger\right).
\end{align}
Defining $\gamma_{\pm}=\frac12\left(\gamma_x\pm i\gamma_y\right),$ the corresponding contribution to the self energy is given by 
\begin{align}
    & \Sigma_{1}+\Sigma_2 = \int_q \frac{\lambda^2}{M_q} \left[6\frac{\chi_{d}^{-1}(q)}{4}\gamma_zG_{0,k+q}\gamma_z+12\chi_N^{-1}(q)\left(\gamma_+G_{0,k+q}\gamma_-+\gamma_-G_{0,k+q}\gamma_+\right)\right]\\
     &=\int_{\bq}T\sum_{i\Omega\in \text{Bosonic}}\frac{\lambda^2}{M_q}\frac{1}{(i\omega+i\Omega)^2-\epsilon_{\bk+\bq}^2}\left[\frac23 (g_d-(i\Omega)^2)\left(i\omega+i\Omega+\epsilon_{\bk+\bq}\gamma_z\right)+ 12(g_N-(i\Omega)^2)(i\omega+i\Omega-\epsilon_{\bk+\bq}\gamma_z)\right].
\end{align}
Note that $\gamma_zG_0\gamma_z=G_0 = i\omega-\epsilon \gamma_z$, while $\gamma_-G_0\gamma_+ + \gamma_+G_0\gamma_-=i\omega+\epsilon\gamma_z$. As a result, if we consider the self energies as function of $i\omega$ and $\epsilon_{\bk+\bq}$, we find that $\Sigma_1\sim \lambda^2\int_q \frac{3\chi_d^{-1}(q)}{2M_q}G_0(i\omega,\epsilon_{\bk+\bq})$ while $\Sigma_2\sim \lambda^2\int_q \frac{12\chi_N^{-1}(q)}{M_q}G_0(i\omega,-\epsilon_{\bk+\bq})$. This allows us to argue the effect of $\Sigma_2$ pushing high energy states towards the vicinity of $\omega=0$, while $\Sigma_1$ pushes states away from $\omega=0$.

To perform the Matsubara sums, we define
\begin{align}
    h(i\omega,\epsilon,g)=&T\sum_{i\Omega}\frac{-(i\Omega)^2+g}{((i\Omega)^2-E_+^2)((i\Omega)^2-E_-^2)}\frac{1}{i\omega+i\Omega-\epsilon}\\
    &=\frac12\frac{1}{E_+^2-E_-^2}\left( \frac{E_+^2-g}{E_+}\left(K(i\omega,\epsilon,E_+)-K(i\omega,\epsilon,-E_+)\right)-\frac{E_-^2-g}{E_-}\left(K(i\omega,\epsilon,E_-)-K(i\omega,\epsilon,-E_-)\right)\right),
\end{align}
with $K(i\omega,\epsilon,E)$ as defined in \eqref{eq:keq}.
In terms of these functions, the self energy is given by
\begin{align}
    \Sigma_1&=\lambda^2\int_{\bq}\frac{1}{3}\left[\left(h(i\omega,\epsilon_{\bk+\bq},g_d)+h(i\omega,-\epsilon_{\bk+\bq},g_d)\right)+\left(h(i\omega,\epsilon_{\bk+\bq},g_d)-h(i\omega,-\epsilon_{\bk+\bq},g_d)\right)\gamma_z\right],\\
\Sigma_2&=\lambda^2\int_{\bq}6\left[\left(h(i\omega,\epsilon_{\bk+\bq},g_N)+h(i\omega,-\epsilon_{\bk+\bq},g_N)\right)-\left(h(i\omega,\epsilon_{\bk+\bq},g_N)-h(i\omega,-\epsilon_{\bk+\bq},g_N)\right)\gamma_z\right].
\end{align}

We can expand the total self energy $\Sigma=\Sigma_1+\Sigma_2+\Sigma_3$ in terms of Pauli matrices in Nambu space,
\begin{align}
    \Sigma(k)=\Sigma_{Id}(k)+\Sigma_{z}(k)\gamma_z+\Sigma_{\gamma_y}(k)\gamma_y,
\end{align}
where
\begin{align}
    \Sigma_{Id}(k)&=\lambda^2\int_{\bq}\left[\frac{1}{3}\left(h(i\omega,\epsilon_{\bk+\bq},g_d)+h(i\omega,-\epsilon_{\bk+\bq},g_d)\right)+6\left(h(i\omega,\epsilon_{\bk+\bq},g_N)+h(i\omega,-\epsilon_{\bk+\bq},g_N)\right)\right],\\
    \Sigma_{z}(k)&=\lambda^2\int_{\bq}\left[\frac{1}{3}\left(h(i\omega,\epsilon_{\bk+\bq},g_d)-h(i\omega,-\epsilon_{\bk+\bq},g_d)\right)-6\left(h(i\omega,\epsilon_{\bk+\bq},g_N)-h(i\omega,-\epsilon_{\bk+\bq},g_N)\right)\right],\\
    \Sigma_{\gamma_y}(k)&=3\phi_0\lambda^2\int_{\bq}\left[f(i\omega,\epsilon_{\bk+\bq})-f(i\omega,-\epsilon_{\bk+\bq})\right].\label{eq:gammay}
\end{align}

\section{Suppression of DOS at $\omega=0$}
\label{sec:suppressiondos}
In this section, we derive a compact approximate analytical expression for the suppression of the density of states (DOS) as a result of the anomalous term $\Sigma_3=\Sigma_{\gamma_y}\gamma_y$. To this end, we focus on the limit of large bosonic velocities $v_\mu$ in $\chi_\mu$ and replace the $\bq$ integral in \equref{eq:gammay} with the value of the integrand at $\bq=0$,
\begin{align}
     \Sigma_{\gamma_y}(\omega+i0^+,\bk)&=3\phi_0\lambda^2\frac{r_N}{v_N^2}\left(f(\omega+i0^+,\epsilon_{\bk})-f(\omega+i0^+,-\epsilon_{\bk})\right).
\end{align}
Note that we would first need to re-parametrize the integral in terms of $\tilde{\bq}=\bq\sqrt{r_N}/v_N$ and then set $\tilde\bq=0$. This approximation would then be valid in the large $v_d/v_N$ limit with this re-scaling. We then Taylor expand $f(z,\epsilon)$ with respect to $\epsilon,\omega$, at a non-zero finite $T$ (satisfying $\epsilon\ll T \ll \sqrt[4]{r_dr_N-\phi_0^2}$). In this limit, we find the self energy to be
\begin{align}
    \Sigma_{\gamma_y}=\frac{3\phi_0r_N\lambda^2}{2v_N^2T(r_dr_N-\phi_0^2)}\epsilon_{\bk} = \alpha\epsilon_{\bk}.
\end{align}
This expression is in agreement with the result in the main text [\figref{fig:DOSFull}(a)] which shows that as $\epsilon\to 0$, the contribution of $\Sigma_y$ vanishes. With such a self-energy, the spectral function is given by 
\begin{align}
   A(\omega) = -\frac{1}{\pi}\text{Im }\frac{\omega+i0^+}{(\omega+i0^+)^2-(1+\alpha^2)\epsilon_{\bk}^2}.
\end{align}
A simple way to look at this, is that the band structure is simply renormalized as $\epsilon_{\bk}\to \sqrt{1+\alpha^2}\epsilon_{\bk} $. This reduces the effective band mass, and thus the DOS is suppressed by a factor of $\sqrt{1+\alpha^2}$, as stated in the main text.
\begin{figure}[tb]
   \centering
    \includegraphics[width=\linewidth]{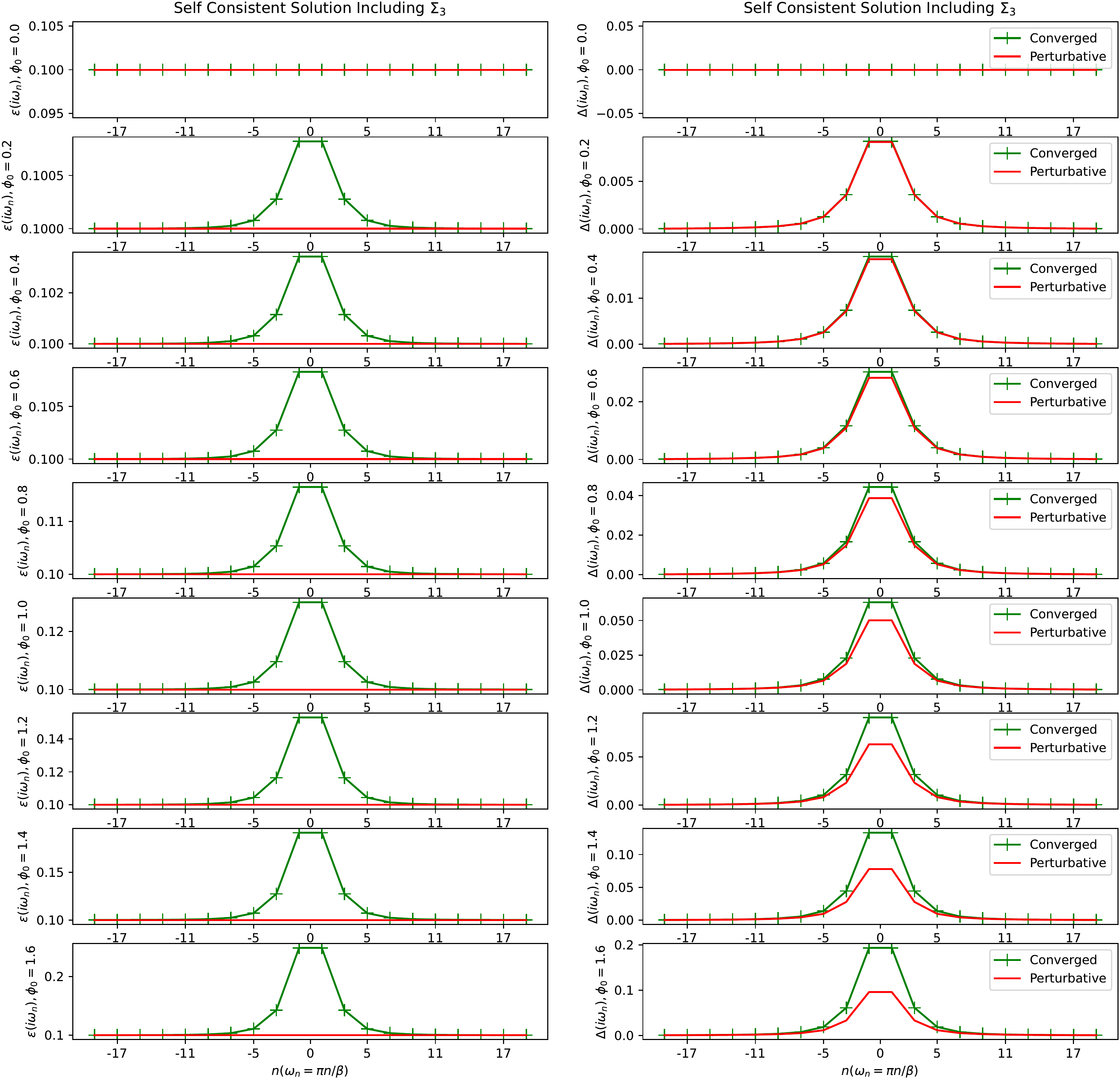}
    \caption{The first order solution to $\varepsilon(i\omega),\Delta(i\omega)$ (red) and the self consistent solution (green) for the self energy in Matsubara space. Note the offset by $0.1$ in the $y$ axis in the left column. We chose $\epsilon_{\vec{k}}=0.1,r_d=9,r_N=1,T=\frac{1}{\beta}=0.2,\lambda=1$ and measured all energies in units of $\sqrt{r_N}$.}
    \label{fig:selfconsistentcompare}
\end{figure}

\begin{figure}[tb]
   \centering
    \includegraphics[width=0.45\linewidth]{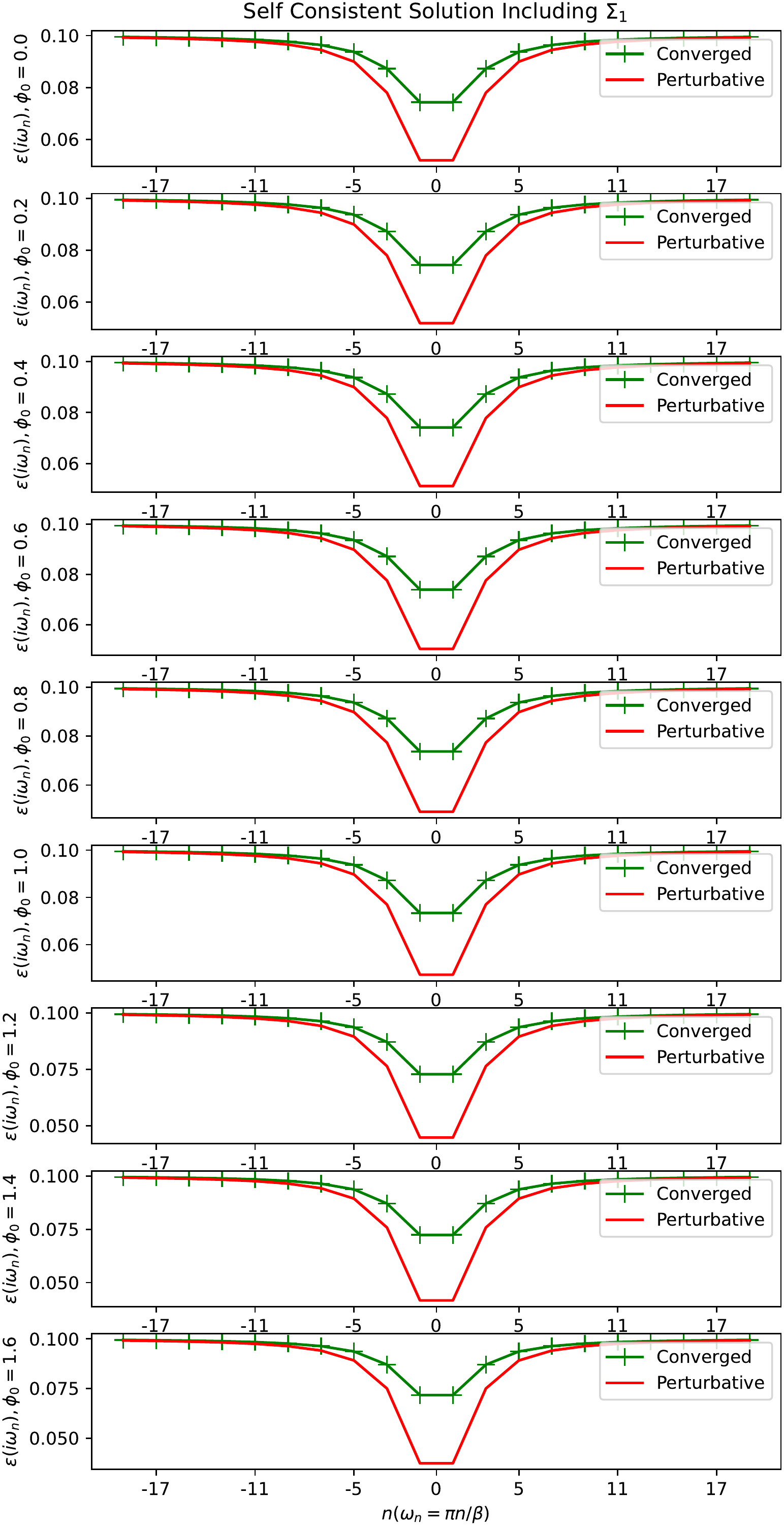}
    \includegraphics[width=0.45\linewidth]{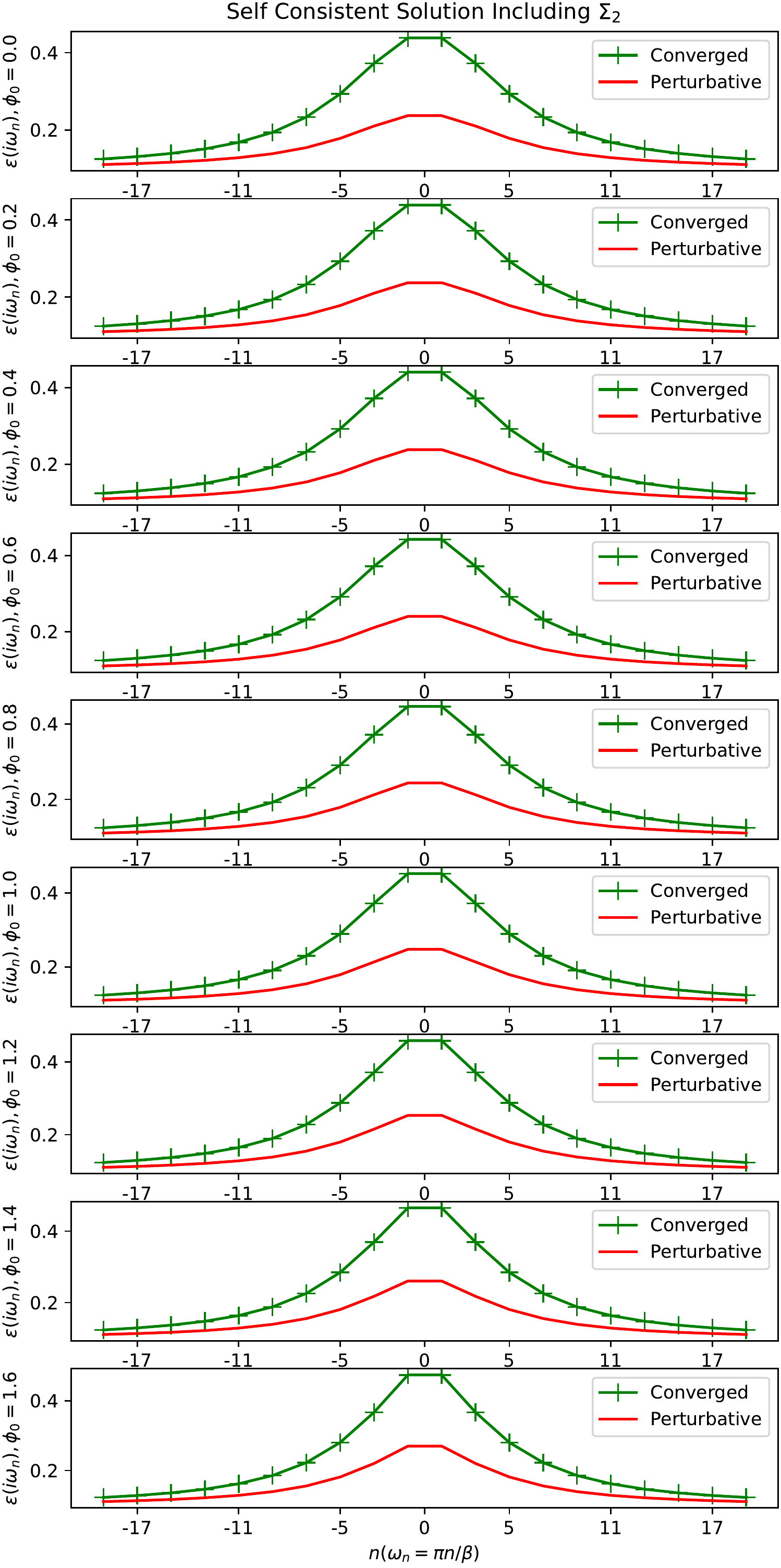}
    \caption{The first order solution to $\varepsilon(i\omega)$ (red) and the self consistent solution (green) after including the effects of $\Sigma_1$ (left column) and $\Sigma_2$ (right column) separately. Same parameters as in \figref{fig:selfconsistentcompare}.}
    \label{fig:selfconsistentcomparesigma21}
\end{figure}

\begin{figure}[tb]
   \centering
    \includegraphics[width=\linewidth]{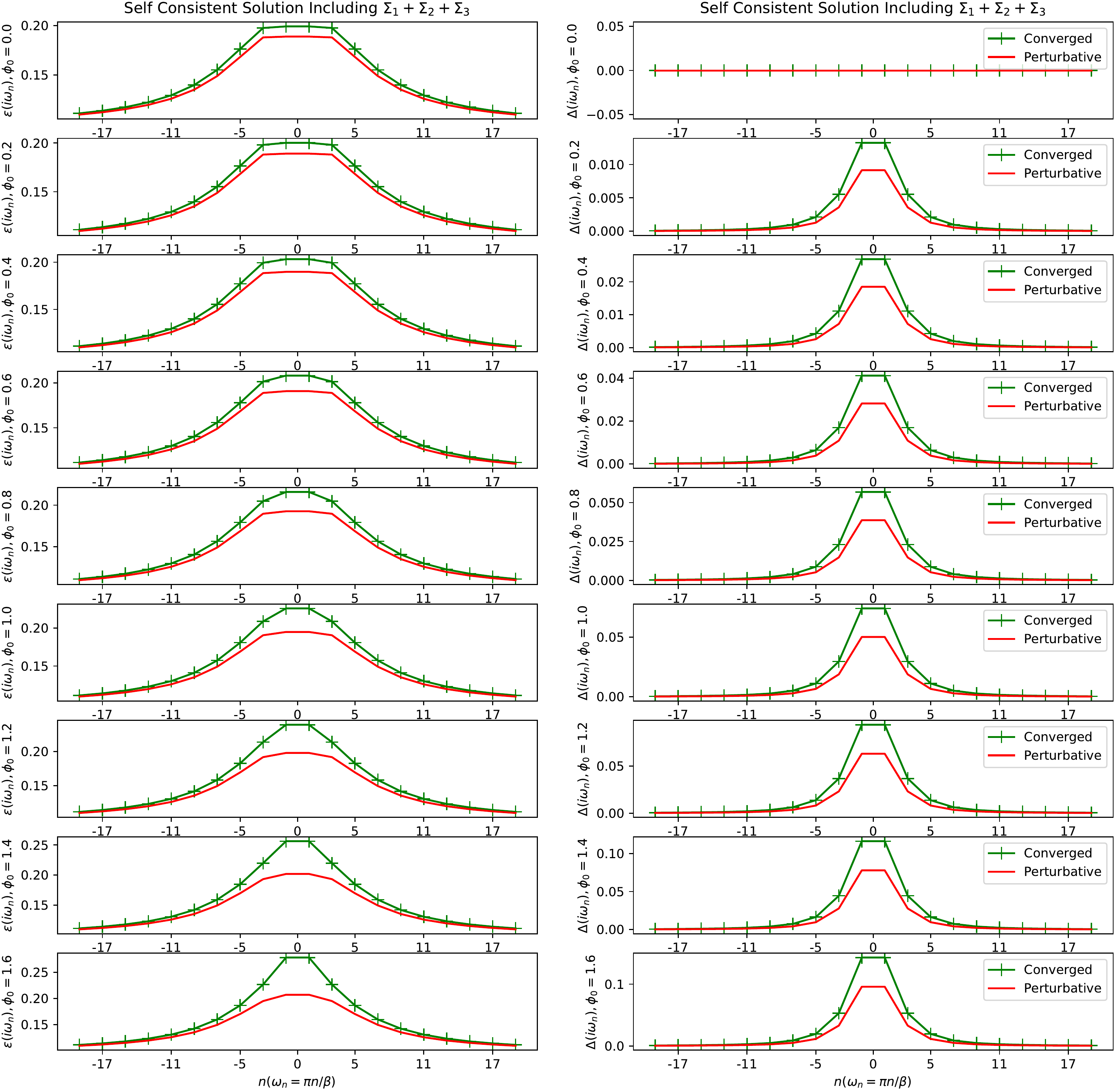}
    \caption{The first order solution to $\varepsilon(i\omega), \Delta(i\omega)$ (red) and the self consistent solution (green) after including the effects of all the terms of the self energy $\Sigma_1+\Sigma_2+\Sigma_3$. Same parameters as in \figref{fig:selfconsistentcompare}.}\label{fig:phidependentselfconsistentfull}
\end{figure} 

\section{Higher-order corrections to electronic Green's function}

In this section, we show comparisons between the first order perturbative solution and the full self consistent solution to the fermionic Green's function. We define the corrected Green's function to be $G(i\omega,\vec{k}) =i\omega Z_{\vec{k}}(i\omega) - \varepsilon_{\vec{k}}(i\omega)\gamma_z+\Delta_{\vec{k}}(i\omega)\gamma_y$. In practice, we find that $Z_{\vec{k}}(i\omega)\approx 1$, so we focus on $\varepsilon_{\vec{k}}(i\omega)$ and $\Delta_{\vec{k}}(i\omega)$ in the following. 

In \figref{fig:selfconsistentcompare}, we show a comparison of the first order result for $\varepsilon_{\vec{k}}(i\omega_n)$ and $\Delta_{\vec{k}}(i\omega_n)$ after including the evaluation of the $\Sigma_3$ term of the self energy  [last diagram in \figref{fig:DOSContr}(b)] and the full self consistent solution to the self energy in Matsubara space [obtained by summing up the diagrams in \figref{fig:DOSContr}(a) corresponding to $\Sigma_3$] at fixed $\vec{k}$. We find that for small values up to $\phi_0\sim 0.6r_N$, the first order and self consistent solutions differ little. In first order, $\varepsilon_{\vec{k}}(i\omega_n)$ does not get renormalized since $\Sigma_3$ acquires a $\gamma_z$ term only if the Green's function has a $\gamma_y$ term. Such a $\gamma_y$ term does not exist in the normal state about which we perform perturbation theory.  As $\phi_0$ increases, we find that the self consistent solution is lower in magnitude that the first order solution. 

In \figref{fig:selfconsistentcomparesigma21}, we show the corrections in $\varepsilon(i\omega_n)$ after including the effects of $\Sigma_1$ (left column) and $\Sigma_2$ (right column). As expected and argued in the main text, we find that $\Sigma_1$ and $\Sigma_2$ have qualitatively the opposite effects on the renormalization of $\varepsilon(i\omega_n)$. In both the cases, we find that the magnitude of the self consistent solution is higher than the perturbative corrections. However, since the fermionic Matsubara frequencies do not contain $0$, we cannot directly say what this implies for the solution 
on the real axis. The magnitude of $\phi_0$ has little effect on the solution since the effect of spin and triplet fluctuations are controlled by $g_N$ and $g_d$, respectively, which we keep constant. 

In \figref{fig:phidependentselfconsistentfull}, we plot the corrections in $\varepsilon(i\omega_n)$ and $\Delta(i\omega_n)$ after including the effects of all the self energies $\Sigma=\Sigma_1+\Sigma_2+\Sigma_3$. We find that the inclusion $\Sigma_1$ and $\Sigma_2$ together reduces the difference between the self consistent and perturbative solution (refer to the plot near $\phi_0\sim 0$). As we increase $\phi_0$, the difference between the self consistent and perturbative solution increases due to the effect of $\Sigma_3$ which is controlled by $\phi_0$. 

Taken together, we see that the inclusion of second- and higher-order diagrams that contribute in the large-$N$ limit defined in the main text yields qualitatively similar behavior on the imaginary axis compared to the first-order diagrams. We therefore expect that the qualitative picture that $\mathcal{S}_1$ renormalizes the DOS close to the Fermi level on top of which $\mathcal{S}_2$ reduces the low-energy spectral weight still applies. Since the impact of $\mathcal{S}_2$ is controlled by small $\phi_0$ and good quantitative agreement is found for $\phi_0$ up to $0.6 r_N$, we expect that \figref{fig:DOSFull}(c) would look similar when higher-order corrections were included. 

\section{Meissner Effect From ODLRO}
The consequences of ODLRO defined in terms of four-fermion or two-boson correlators are well-known \cite{yangConceptOffDiagonalLongRange1962,sewellOffdiagonalLongrangeOrder1990,niehOffdiagonalLongrangeOrder1995,sewellOffdiagonalLongRange1997}. As a result of spin-rotation symmetry, we cannot capture ODLRO using a correlator of only two bosons. Instead, we have to study the four-boson density matrix
\begin{align}
    \rho(\bx_1,\bx_2,\bx_1',\bx_2') = \langle \bN(\bx_1)\cdot\bd^*(\bx_2) \bN(\bx_1')\cdot\bd(\bx_2')\rangle.
\end{align}
Although the derivation is in close analogy to the two-boson or four-fermion case, we here show explicitly how the Meissner effect follows from 
\begin{equation}
    \rho(\bx_1,\bx_2,\bx_1',\bx_2') \to \phi_0^*(\bx_1,\bx_2)\phi_0(\bx_1',\bx_2') \neq 0, \quad |\vec{x}_j-\vec{x}_j'|\rightarrow \infty. \label{FourBosonODLRO}
\end{equation}
Let us consider the system to be in the presence of a spatially uniform orbital magnetic field of strength $\bB=B_0\hat{z}$ in the out of plane direction. Note that an in-plane orbital magnetic field does not couple to the bosons as the spatial motion is constrained to the two-dimensional plane of the system. The corresponding vector potential is given by $\bA(\bx)= \frac12 \bB\times \bx$, with $\vec{x}=(x,y,0)$. Under an in-plane translation by $\ba$, the vector potential transforms as 
\begin{align}
    \bA(\bx)\to \bA(\bx-\ba)& =  \bA(\bx) -\frac12 \bB\times \ba \\
    & =  \bA(\bx) -\frac12\nabla \left[ \ba\cdot(\bx\times \bB)\right]\\
    &=\bA(\bx) +\nabla\chi_{\vec{a}}(\bx),
\end{align}
where $\chi_{\vec{a}}(\bx)=-\frac12\ba\cdot(\bx\times \bB)$. Note that the triplet pairing field $\bd$ is a charge-$2e$ bosonic field, while the magnetization field $\bN$ is neutral. Therefore, under simultaneous gauge transformation and displacement by $\vec{a}$ in the presence of a magnetic field, the fields transform as 
\begin{align}
    \bd(\bx) &\rightarrow e^{i\frac{2e}{\hbar c}\chi_a(\bx)}\bd(\bx-\vec{a}),\\
    \bN(\bx) &\rightarrow \bN(\bx-\vec{a}), \\
    \vec{A}(\vec{x}) &\rightarrow  \vec{A}(\vec{x}).
\end{align}
As a result of gauge covariance and translational symmetry, the four-body density matrix obeys
\begin{align}
    \rho(\bx_1,\bx_2,\bx_1',\bx_2') =  e^{i\frac{2e}{\hbar c}\left(\chi_{\vec{a}}(\bx_2')-\chi_{\vec{a}}(\bx_2)\right)}\rho(\bx_1-\ba,\bx_2-\ba,\bx_1'-\ba,\bx_2'-\ba). \label{GaugeCovarianceCondTransl}
\end{align}

Now suppose the system has ODLRO, i.e., \equref{FourBosonODLRO} holds. In combination with \equref{GaugeCovarianceCondTransl}, this implies
\begin{align}
   \phi_0^*(\bx_1,\bx_2)\phi_0(\bx_1',\bx_2') &=  e^{i\frac{2e}{\hbar c}\left(\chi_{\vec{a}}(\bx_2')-\chi_{\vec{a}}(\bx_2)\right)}\phi_0^*(\bx_1-\ba,\bx_2-\ba)\phi_0(\bx_1'-\ba,\bx_2'-\ba)\\
   \Longrightarrow \phi_0(\bx_1,\bx_2)&=f_{\ba}e^{i\frac{2e}{\hbar c}\chi_{\vec{a}}(\bx_2)}\phi_0(\bx_1-\ba,\bx_2-\ba),
\end{align}
where $f_{\ba}$ is a position-independent phase factor. Now suppose we perform two different translations by $\ba$ and $\bb$. We can perform $\ba$ first and then $\bb$. Alternatively, we can do $\bb$ first and then $\ba$. They respectively give us
\begin{align}
    \phi_0(\bx_1,\bx_2)&=f_{\bb}f_{\ba}e^{i\frac{2e}{\hbar c}\chi_{\vec{a}}(\bx_2)}e^{i\frac{2e}{\hbar c}\chi_{\vec{b}}(\bx_2-\ba)}\phi_0(\bx_1-\ba,\bx_2-\ba),\\
    \phi_0(\bx_1,\bx_2)&=f_{\bb}f_{\ba}e^{i\frac{2e}{\hbar c}\chi_{\vec{b}}(\bx_2)}e^{i\frac{2e}{\hbar c}\chi_{\vec{a}}(\bx_2-\bb)}\phi_0(\bx_1-\ba,\bx_2-\ba).
\end{align}
This is only consistent if
\begin{align}
    e^{i\frac{2e}{\hbar c}\left(\chi_{\vec{b}}(\bx_2)+\chi_{\vec{a}}(\bx_2-\bb)-\chi_{\vec{a}}(\bx_2)-\chi_{\vec{b}}(\bx_2-\ba)\right)}=1.
\end{align}
We can evaluate $\chi_{\vec{b}}(\bx_2)+\chi_{\vec{a}}(\bx_2-\bb)-\chi_{\vec{a}}(\bx_2)-\chi_{\vec{b}}(\bx_2-\ba)=\bB\cdot\left(\ba\times \bb\right)$. Thus, the condition for equality of phases becomes
\begin{align}
    \frac{2e}{\hbar c}\bB\cdot \left(\ba\times \bb\right)=2\pi n,
\end{align}
for some integer $n$. The only solution for arbitrary $\ba,\bb$ is thus $\bB=0$.

\section{Demonstration of Off Diagonal Long Range Order}
In this section, we calculate the ODLRO wavefunctions for both the bosons and fermions. The idea is to calculate the $4-$body correlator $\langle \bN(\bx'_1)\cdot \bd(\bx_2')^* \bN(\bx_1)\cdot \bd(\bx_2) \rangle$ for the bosons and $\langle c^\dagger_{\tau_1's_1'}(\bx_1')c^\dagger_{\tau_2's_2'}(\bx_2')c_{\tau_1,s_1}(\bx_1)c_{\tau_2,s_2}(\bx_2)\rangle$ for the fermions. Due to the $U(1)$ symmetry breaking mediated by $\bN\cdot \bd$ attaining a finite expectation value (and correspondingly $c^\dagger_\tau c^\dagger_{-\tau}$ for the fermions), the ODLRO factorizes into a product of functions of $\bx_1-\bx_2$ and $\bx_1'-\bx_2'$ in the limit $\bx-\bx'\to\infty$, where $\bx=\frac{\bx_1+\bx_2}{2}$ and $\bx'=\frac{\bx_1'+\bx_2'}{2}$, giving rise to ODLRO.   
These wavefunctions decay as a function of their respective relative coordinates $\bx_1-\bx_2$ and $\bx_1'-\bx_2'$. We now calculate these ``macroscopic wavefunctions'' explicitly for the bosonic and fermionic cases.

\subsection{Bosonic ODLRO}
The bosonic ODLRO is given by $\langle \bN(\bx'_1)\cdot \bd(\bx_2')^* \bN(\bx_1)\cdot \bd(\bx_2) \rangle\approx  \langle \bN(\bx'_1)\cdot \bd(\bx'_2)^*\rangle\langle \bN(\bx_1)\cdot \bd(\bx_2) \rangle$ as $\bx-\bx'\to\infty$. All the correlators are evaluated at time $t=0$. As discussed in the main text, to demonstrate ODLRO, it is sufficient to evaluate these correlators to first non-trivial order in the coupling constants. For bosonic ODLRO it is in fact sufficient to focus on zeroth order, i.e., neglecting the coupling to the fermions. Using the translation invariance of the system (and summing over the Matsubara frequencies $i\Omega$ since we are evaluating the correlator at time $t=0$), we then have
\begin{align}
    \psi_B(\bx)&=\langle \bN(\bx)\cdot \bd(\bx=0) \rangle = \int_{\bq}T\sum_{i\Omega}e^{i\bq\cdot \bx}\langle \bN_{-q}\cdot \bd_q\rangle\\
    &= \int_{\bq}T\sum_{i\Omega}e^{i\bq\cdot r}\frac{\phi_0}{[(i\Omega)^2-E_+^2(\vec{q})][(i\Omega)^2-E_-^2(\vec{q})]}\\
    &= \int_{\bq} e^{i\bq\cdot \bx}\frac{\phi_0}{2E_+(\vec{q})E_-(\vec{q})(E_+(\vec{q})+E_-(\vec{q}))} \label{qIntegration}\\
    &\approx \int_{\bq} e^{i\bq\cdot \bx}\frac{\phi_0}{a+b\bq^2}\\
    &=\frac{\phi_0}{b} \int_{\bq}e^{i\bq\cdot \sqrt{\frac{a}{b}}\bx}\frac{1}{1+\bq^2}=2\pi \phi_0 K_0\left(\sqrt{\frac{a}{b}}|\bx|\right)/b\\
    &=2\pi \phi_0 K_0\left(|\bx|/\xi\right)/b, \label{FinalAsymptoticForm}
\end{align}
where $K_0$ is the zeroth modified Bessel function of second kind. In the third line, we evaluated the Matsubara sum at $T=0$, and in the fourth line we series expanded $2E_+(\bq)E_-(\bq)\left(E_+(\bq)+E_-(\bq)\right)$ about $\bq=0$ up to quadratic order. The length scale $\xi=\sqrt{\frac{b}{a}}$ is determined by $r_\mu,v_\mu$. In the limit of $|v_N-v_d|\ll v_N+v_d$, we get
\begin{align}
    \xi=\frac{1}{2} \sqrt{\frac{({v_N^2}+{v_d^2}) \left(\sqrt{{r_N} {r_d}-\phi_0^2} +{r_N}+{r_d}\right)}{{r_N}{r_d}-\phi_0^2}}.
\end{align}

In \figref{fig:ODLRO}(b), we plot the numerical ODLRO wavefunction $\psi_B(\bx)$ with the full functional dependence on $\bq$ in \equref{qIntegration} included, and compare it with the asymptotic analytical form in \equref{FinalAsymptoticForm}. We find good agreement between the numerical and analytical results.

\subsection{Fermionic ODLRO}
Similarly, we can find the fermionic ODLRO, which in real space is generically written as $\langle c^\dagger_{\tau_1's_1'}(\bx_1')c^\dagger_{\tau_2's_2'}(\bx_2')c_{\tau_1,s_1}(\bx_1)c_{\tau_2,s_2}(\bx_2)\rangle\sim \langle c^\dagger_{\tau_1's_1'}(\bx_1')c^\dagger_{\tau_2's_2'}(\bx_2')\rangle\langle c_{\tau_1,s_1}(\bx_1)c_{\tau_2,s_2}(\bx_2)\rangle$ in the limit $\bx_j-\bx'_j\to\infty$. Here, $\tau,s$ are the valley and spin indices respectively. To demonstrate ODLRO, we thus have to evaluate the $2-$fermion correlators, which in momentum space becomes
\begin{equation}
    \left(\Psi^*_\text{F}(\vec{x})\right)_{s_1,s_2}=\langle c^\dagger_{\tau_1,s_1}(\bx,t=0)c^\dagger_{\tau_2,s_2}(\bx=0,t=0)\rangle = \int_k e^{i\bk\cdot\bx}\langle c^\dagger_{k,\tau_1,s_1}c^\dagger_{-k,\tau_2,s_2}\rangle. \label{ExpressionForFermionicODLRO}
\end{equation}
Since the superconducting pairing takes place only between electrons between opposite valleys, we will have only $\tau_2=-\tau_1$ giving non-zero correlators. Without loss of generality we chose $\tau_1=+, \tau_2=-$. 
 Up to first order in $\phi_0$, we have
\begin{align}
    \langle c^\dagger_{k,+,s_1}c^\dagger_{-k,-,s_2}\rangle&=\langle c^\dagger_{k,+,s_1}c^\dagger_{-k,-,s_2}\left(-\int_q\frac12\frac{\phi_0\lambda^2}{M_q}\bS_{q}\cdot\bD_{q}^\dagger\right)\rangle_0, \label{FermionicCorrelator}
\end{align}
where $\langle ...\rangle$ is the average with respect to the interacting and $\langle...\rangle_0$ with respect to the non-interacting ground state. We define $G(k)=\delta_{ss'}\delta_{\tau\tau'}G_{V,k}=\frac{\delta_{ss'}\delta_{\tau\tau'}}{i\omega_n-\epsilon_{\bk}}=-\langle c_{s,\tau}c^\dagger_{s',\tau'}\rangle$ to be the Green's function in the fermionic basis (assuming $\epsilon_{\vec{k}}=\epsilon_{-\vec{k}}$).  Equation~(\ref{FermionicCorrelator}) can then be evaluated as,
\begin{align}
   -\frac{\phi_0\lambda^2}{2}&\int_q\frac{1}{M_q}  \langle c^\dagger_{k,+,s_1}c^\dagger_{-k,-,s_2}\left(\bS_{q}\cdot\bD_{q}^\dagger\right)\rangle_0\\
   &=-\frac{\phi_0\lambda^2}{2}\int_q\frac{1}{M_q}\langle   c_{k+,s_1}^\dagger c_{-k-,s_2}^\dagger\left(\sum_{k_1,k_2,p1=\pm,p_2=\pm}p_1p_2 \left(c^\dagger_{k_1+q,p_1} \bs c_{k_1,p_1}\right)\cdot \left(c_{k_2+q,p_2}is_y\bs c_{-k_2,-p_2}\right)\right)\rangle_0\\
    &=-2\frac{\phi_0\lambda^2}{2}\int_q\frac{1}{M_q}\langle   c_{k+,s_1}^\dagger c_{-k-,s_2}^\dagger\left(\sum_{k_1,p}- \left(c_{-k_1,-p}is_y\bs(- G_{V,k_1+q,p} )\bs c_{k_1,p}\right)\right)\rangle_0\\
    &=-6\frac{\phi_0\lambda^2}{2}\int_q\frac{1}{M_q}\langle   c_{k+,s_1}^\dagger c_{-k-,s_2}^\dagger\left(\sum_{k_1,p}\left(c_{-k_1,-p}is_y G_{V,k_1+q,p}  c_{k_1,p}\right)\right)\rangle_0\\
    &=-6\frac{\phi_0\lambda^2}{2}\int_q\frac{1}{M_q}\langle   c_{k+,s_1}^\dagger c_{-k-,s_2}^\dagger\left(\sum_{k_1} \left(c_{-k_1,-}is_y G_{V,k_1+q,+}  c_{k_1,+}+c_{-k_1,+}is_y G_{V,k_1+q,-}  c_{k_1,-}\right)\right)\rangle_0\\
    &=-6\frac{\phi_0\lambda^2}{2}\int_q\frac{1}{M_q} \left(-(-G_{V,-k,-})(is_y)_{s_2s_1} G_{V,k+q,+} G_{V,k,+}+(-G_{V,k,+})(is_y)_{s_1s_2} G_{V,-k+q,-}  G_{V,-k,-}\right)\\
    &=-6 \frac{\phi_0\lambda^2}{2}\int_q\frac{1}{M_q}G_{V,k}G_{V,-k}\left(G_{V,-k+q}+G_{V,k+q}\right)(is_y)_{s_2s_1}.
\end{align}
We continue by calculating the Matsubara sum over $i\Omega_n$ and over $i\omega_n$ [see \equref{ExpressionForFermionicODLRO}],
\begin{align}
   T^2\sum_{i\omega_n,i\Omega_n}&\frac{1}{(i\Omega_n^2-E_+(\bq)^2)(i\Omega_n^2-E_-(\bq)^2)} G_{V,k}G_{V,-k}\left(G_{V,-k+q}+G_{V,k+q}\right)\\
   &=-T^2\sum_{i\omega_n,i\Omega}\frac{1}{(i\omega_n)^2-\epsilon_{\bk}^2}\frac{1}{((i\Omega_n)^2-E_+(\bq)^2)((i\Omega_n)^2-E_-(\bq)^2)}\left(\frac{1}{i\omega_n+i\Omega_n-\epsilon_{-\bk+\bq}}+\frac{1}{-i\omega_n+i\Omega_n-\epsilon_{\bk+\bq}}\right)\\
   &=:X(\epsilon_{\bk},\bq).
\end{align}
For simplicity, we here focus on the limit where the remaining sum over $\vec{q}$ in \equref{FermionicCorrelator} is determined by its $\vec{q}=0$ component. With $E_\pm \equiv E_\pm(\bq=0)$ and $v_N,r_N=1$, we have
\begin{align}
    &\hat{X}(\epsilon)\equiv X(\epsilon,\bq\to 0) \\
    &=\frac{n_f(\epsilon)^2}{2\epsilon}\left(-2\frac{e^{\beta \epsilon}}{ E_+^2E_-^2}+\frac{2}{\left(E_+^2-4\epsilon^2\right)\left(E_-^2-4\epsilon^2\right)}+\left(\frac{2\epsilon n_B(E_+)-E_+n_f(E_+)}{E_+(E_+^2-E_-^2)(E_+^2-4\epsilon^2)n_f(E_+)n_B(2\epsilon)}+E_+\leftrightarrow E_-\right)\right),
\end{align}
we can then finally write
\begin{align}
   \Psi^*_\text{F}(\vec{x}) &=3|\phi_0|\lambda^2s_{y}\left(\frac{1}{V}\sum_{\vec{k}} e^{i{\bk}\cdot \bx} \hat{X}(\epsilon_{\bk})\right), \\
   & = \frac{3|\phi_0|\lambda^2s_{y}}{2\pi}\int_0^\infty dk  k J_0({\bk}\cdot \bx)\hat{X}(\hbar^2({\bk}^2-\bk_F^2)/(2m)).
\end{align}
In the second line, we assumed $\epsilon_{\bk} = \hbar^2({\bk}^2-{\bk}_F^2)/2m$.
Using this expression, we calculate the spatial profile of the fermionic ODLRO wavefunction  numerically for various values of $\epsilon_F\equiv\epsilon_{\bk_F}$ in \figref{fig:ODLRO}(a). Unlike the case of the bosonic ODLRO (which was exponentially decaying), the fermionic ODLRO has an oscillating component superimposed on an exponentially decaying envelope.

\section{Ginzburg-Landau theory}
We here calculate the Landau-Ginzburg theory for the bosonic superfluid condensate parameter to leading (zeroth) order in the fermion-boson coupling $\lambda$. To tis end, we assume that $\phi_0$ is now spatially and temporally varying. This results in non-zero Fourier modes $\phi_{q}$ for $\bq,i\Omega\ne 0$. 

In momentum space, the bosonic action is generalized according to
\begin{align}
    \mathcal{S}_{B} &= \int_q[\chi_N^{-1}(q) \bN_{q}\cdot\bN_{-q} + \chi_{SC}^{-1}(q)\bd_{q}^*\cdot\bd_{q}+(\phi_0 \bd_{q}\cdot \bN_{-q} + \text{H.c.})]\\
    &=\int_q \begin{pmatrix}\bN_{-q}^T& {\bd_q^\dagger}\end{pmatrix}\begin{pmatrix}\chi_N^{-1}(q)&\phi_0\\\phi_0&\chi_d^{-1}(q)\end{pmatrix}\begin{pmatrix}\bN_{q}\\ \bd_q\end{pmatrix}\\
    &\to\int_{q,k} \begin{pmatrix}\bN_{-q-q_2}^T& {\bd_{q+q_2}^\dagger}\end{pmatrix}\begin{pmatrix}\chi_N^{-1}(q)\delta_{q_2=0}&\phi_{q_2}\\\phi_{-q_2}^*&\chi_d^{-1}(q)\delta_{q_2=0}\end{pmatrix}\begin{pmatrix}\bN_{q}\\ \bd_q\end{pmatrix}
\end{align}
So after integrating out $\vec{d}$ and $\vec{N}$, the effective action for $\phi$ reads as
\begin{equation}
    \mathcal{S}_{\text{eff}} = \frac12 Tr\ln G^{-1}[\phi],
\end{equation}
where 
\begin{align}
   G^{-1}[\phi](q+q_1,q)&=G_0^{-1}(q)\delta_{q_1,0}+\Gamma_{q+q_1,q}\\
   G_0^{-1}&=\begin{pmatrix}\chi_N^{-1}(q)&0\\0&\chi_d^{-1}(q)\end{pmatrix}\\
   \Gamma_{q+q_1,q}&=\begin{pmatrix}0&\phi_{q_1}\\\phi_{-q_1}^*&0\end{pmatrix}.
\end{align}
To derive the Ginzburg-Landau theory for $\phi$, we expand $Tr\ln G^{-1}$ upto second order in $\Gamma$, which is equivalent to second order in $\phi$. This gives us
\begin{align}
    S_{\text{GL}}= Tr\ln (G_0^{-1}+\Gamma) \approx Tr G_0^{-1} +Tr G_0\Gamma-\frac12Tr G_0\Gamma G_0\Gamma 
\end{align}
Because of the diagonal structure of $G_0$, and the off diagonal structure of $\Gamma$, the linear term $TrG_0\Gamma$ is $0$. The quadratic term becomes 
\begin{align}
    \sum_{q',q}Tr G_0(q'+q)\Gamma(q'+q,q')G_0(q')\Gamma(q',q'+q)  &= \sum_{q',q}Tr  \begin{pmatrix}0&\chi_N(q'+q)\phi_{q}\\\chi_d(q'+q)\phi_{-q}^*&0\end{pmatrix}\begin{pmatrix}0&\chi_N(q')\phi_{-q}\\\chi_d(q')\phi_{q}^*&0\end{pmatrix}\\
    &=\sum_{q',q} \chi_N(q'+q)\chi_d(q')\phi_{q}\phi_{q}^* + \chi_N(q')\chi_d(q'+q)\phi_{-q}\phi_{-q}^*\\
    &=\sum_{q',q}\left( \chi_N(q'+q)\chi_d(q') + \chi_N(q')\chi_d(q'-q)\right)\phi_{q}\phi_{q}^*\\
    &=\sum_{q',q}\left( \chi_N(q'+q)\chi_d(q') + \chi_N(q'+q)\chi_d(q')\right)\phi_{q}\phi_{q}^*\\
    &=2\sum_{q',q} \chi_N(q'+q)\chi_d(q')\phi_{q}\phi_{q}^*
\end{align}
We need to evaluate
\begin{align}
    &\sum_{q'}  \chi_N(q'+q)\chi_d(q') = \int_{\bq'}T\sum_{i\Omega'\in \text{Bosonic}} \left(\frac{1}{((i\Omega'+i\Omega)^2-r_N-v_N^2(\bq'+\bq)^2)((i\Omega')^2-r_d-v_d^2 \bq'^2)}\right)\\
    &= -\frac12\int_{\bq'}\left(\frac{1}{\sqrt{r_N+v_N^2 ({\bq'}+{\bq}/2)^2}}+\frac{1}{\sqrt{r_d+v_d^2 ({\bq'}-{\bq}/2)^2}}\right)\left(\frac{1}{i\Omega^2-\left(\sqrt{r_N+v_N^2 ({\bq'}+{\bq}/2)^2}+\sqrt{r_d+v_d^2 ({\bq'}-{\bq}/2)^2}\right)^2}\right).
\end{align} 

By expanding the above expression up to second order in $i\Omega$, $\bq$, we find the effective action for the $\phi$ field to be
\begin{align}
    T\sum_{i\Omega,\bq}\left(r_{\phi}-\rho(i\Omega)^2+v^2\bq^2\right)|\phi_{(\bq,i\Omega)}|^2
\end{align}
where the coefficients are given by
\begin{align}
    r_{\phi}&=-\int_{\bq'}\frac{\pi }{\sqrt{g_d} \sqrt{g_N} \left(\sqrt{g_d}+\sqrt{g_N}\right)}\\
    \rho&=\int_{\bq'}\frac{\pi }{\sqrt{g_d} \sqrt{g_N} \left(\sqrt{g_d}+\sqrt{g_N}\right)^3}\\
    v^2&=\int_{\bq'}\frac{\pi  \left(4 \bq'^2 (\sqrt{g_d}+\sqrt{g_N})\left(\frac{v_d^2}{g_d^{3/2}}-\frac{v_N^2}{g_N^{3/2}}\right) \left(\frac{v_N^2}{\sqrt{g_N}}-\frac{v_d^2}{\sqrt{g_d}}\right)-\left(\sqrt{g_d}+\sqrt{g_N}\right)^2\left( \frac{v_d^2 \left( 3v_d^2 \bq'^2-2 g_d\right)}{g_d^{5/2}}+\frac{v_N^2 \left( 3v_N^2 \bq'^2-2 g_N\right)}{g_N^{5/2}}\right)\right)}{16 \left(\sqrt{g_d}+\sqrt{g_N}\right)^4}\\
    &-\frac{2\pi   \left(\frac{1}{\sqrt{g_d}}+\frac{1}{\sqrt{g_N}}\right) \left(\frac{3 \bq'^2 \left(\sqrt{g_N} v_d^2-\sqrt{g_d} v_N^2\right)^2}{g_d g_N}-\left(\sqrt{g_d}+\sqrt{g_N}\right) \left(\frac{v_d^2 \left( 2 g_d-\bq'^2 v_d^2\right)}{g_d^{3/2}}+\frac{v_N^2 \left(2 g_N-\bq'^2 v_N^2\right)}{g_N^{3/2}}\right)\right)}{16 \left(\sqrt{g_d}+\sqrt{g_N}\right)^4}
\end{align}
with $g_\mu=r_\mu+v_\mu^2\bq'^2$. We numerically calculate the quantities $r_{\phi},\rho,v^2$ and plot it in \figref{fig:ODLRO}(c,d) of the main text.

\section{Self-consistent equations in special limits}
In this appendix, we complement the previous analysis by studying two simple limits of the model for phase (B)---mean-field theory and the limit of zero energy-momentum transfer of the bosons. This allows us to study possible non-perturbative solutions systematically. In both cases, we find that the soft gap behavior obtained within perturbation theory is also found in these descriptions as long as $T$ is large enough/the coupling constants, $\lambda$ or $\phi_0$, are small enough. 

\subsection{Mean-field Theory}

In this section, we consider the effective interaction contributed by the $\mathcal{S}_2$ part of the action between the electrons at time $t=0$, in the limit where we replace the $q$ integral with the corresponding value of the integrand at $q=0$, and then perform a mean-field decomposition of the interaction. Defining the Bogoliubov-de Gennes basis as before, $\xi_k=\begin{pmatrix}c_{k,+}& is_yc_{-k,-}^\dagger\end{pmatrix}^T$, with Pauli matrix $\gamma_i$ acting on it, and $\tilde\phi_0=\phi_0\lambda^2r_N/v_N^2$ the corresponding interaction potential is given by 
\begin{align}
    V&= -\frac12\frac{1}{\chi_{d}^{-1}\chi_N^{-1}-|\phi_0|^2}\left(\tilde\phi_0\bS_{q=0}\cdot \bD_{q=0}^\dagger+ \tilde\phi_0^* \bD_{q=0}\cdot\bS_{-q=0}\right)|_{q=0}\\
    &= -\frac12\frac{1}{r_Nr_d-|\phi_0|^2}\int_{\bk_1,\bk_2}\left[-\tilde\phi_0\left( c^\dagger_{\bk_1} \bs \tau_zc_{\bk_1}\right)\cdot \left(c_{\bk_2} \bs is_y\tau_yc_{-\bk_2}\right)+h.c\right]\\
    &= -\frac{1}{r_Nr_d-|\phi_0|^2}\int_{\bk_1,\bk_2}\left[\tilde\phi_0\left( \xi^\dagger_{\bk_1} \bs\gamma_z\xi_{\bk_1}\right)\cdot \left(\xi_{\bk_2}^\dagger \bs i\gamma_{-}\xi_{\bk_2}\right)+h.c\right],
\end{align}
while the free Hamiltonian is given by 
\begin{align}
    H_0 = \int_{\bk} \xi_{\bk}^\dagger \epsilon_{\bk}\gamma_z\xi_{\bk}.
\end{align}
We consider only the effective Hamiltonian at time $t=0$, which is why there are no Matsuabra indices. 

We perform a Hartree-Fock decomposition of $V$, which gives us 
\begin{align}
    V&= \frac{1}{r_Nr_d-|\phi_0|^2}\int_{\bk_1,\bk_2}\left[\tilde\phi_0\left( \xi^\dagger_{\bk_1} \bs\gamma_z\xi_{\bk_1}\right)\cdot \left(\xi_{\bk_2}^\dagger \bs i\gamma_{-}\xi_{\bk_2}\right)+h.c\right]\\
    &\to \frac{c}{2} \int_{\bk}\xi^\dagger_{\bk}\left(\gamma_yC_{\bk}\gamma_z+\gamma_zC_{\bk}\gamma_y\right)\xi_{\bk},
\end{align}
where $C_{\bk} =- \langle \xi_{\bk}\xi_{\bk}^\dagger\rangle$, $c=6 \frac{\tilde\phi_0}{r_Nr_d-\phi_0^2}$, choosing a gauge with real $\phi_0$; further take $\phi_0$ to be positive such that $c>0$. Note that this correlator is related to the Green's function $G$ by $C_{\bk} =T\sum_{i\omega_n} G(k)$. Note that all the Hartree terms vanish since we do not allow for spontaneous breaking of spin-rotation invariance (recall we study finite $T$ in 2D). The effective $2-$particle Hamiltonian is given by
\begin{align}
    H&=\int_{\bk}\xi_{\bk}^\dagger\left(\epsilon_{\bk} \gamma_z +\frac{c}{2}\gamma_yC_{\bk}\gamma_z+\frac{c}{2}\gamma_zC_{\bk}\gamma_y\right)\xi_{\bk}\label{eq:scfock}\\
    &=\int_{\bk}\xi_{\bk}^\dagger\left[ \tilde{\epsilon}_{\bk}\gamma_z+\tilde\Delta_{\bk}\gamma_y\right]\xi_{\bk}
\end{align}
where $\tilde\epsilon_{\vec{k}},\tilde\Delta_{\vec{k}}$ are the self consistent band structure and gap. Making connection with the diagrammatic self consistency relationship to be discussed below, we can foresee that the resulting self consistent equation we get will be the same as \eqref{eq:scdelta} but with $\tilde\epsilon,\tilde\Delta$ replaced with the corresponding $i\omega_n$ averaged value, and the whole equation itself will be $i\omega_n$ averaged.

The correlators in terms of $\tilde\epsilon,\tilde\Delta$ are given by 
\begin{align}
    C_{\bk} = T\sum_{i\omega_n}\frac{1}{i\omega_n-\left[\tilde{\epsilon}_{\bk}\gamma_z +\tilde{\Delta}_{\bk}\gamma_y\right]} = \frac{n_f(E_{\bk})-n_f(-E_{\bk})}{2E_{\bk}}\left[\tilde{\epsilon}_{\bk}\gamma_z +\tilde{\Delta}_{\bk}\gamma_y\right],
\end{align}
where $E_{\bk}=\sqrt{\tilde\epsilon_{\bk}^2+\tilde\Delta_{\bk}^2}>0$. Thus, using \eqref{eq:scfock}, the self consistency equations become
\begin{align}
    \tilde{\epsilon}_{\bk} &=\epsilon_{\bk} +c\tilde{\Delta}_{\bk}\frac{n_f(E_{\bk})-n_f(-E_{\bk})}{2E_{\bk}}\\
    \tilde\Delta_{\bk}&=c\tilde{\epsilon}_{\bk}\frac{n_f(E_{\bk})-n_f(-E_{\bk})}{2E_{\bk}}. \label{SecondMFSelfConEq}
\end{align}
Let us define $\beta_{\bk}=c\frac{n_f(-E_{\bk})-n_f(E_{\bk})}{2E_{\bk}} = c \frac{\tanh\left(\frac{E_{\bk}}{2T}\right)}{2E_{\bk}}$ and first assume $\beta_{\bk} < 1$, which always holds as long as $T>c/4$. The self consistency equations can then be rearranged as
\begin{subequations}
\begin{align}
    \tilde{\epsilon}_{\bk}&=\frac{1}{1-\beta_{\bk}^2}\epsilon_{\bk}\\
    \tilde\Delta_{\bk}&=\frac{-\beta_{\bk}}{1-\beta_{\bk}^2}\epsilon_{\bk}.
\end{align}\label{SelfConsistEquMF}\end{subequations}
Using this, we find $E_{\bk} = \frac{\sqrt{1+\beta_{\bk}^2}}{1-\beta_{\bk}^2}\epsilon_{\bk}$. Note, however, that $\beta_{\vec{k}}$ also depends on $E_{\vec{k}}$ and, thus, this relation should be thought of as a self consistency equation, to be solved for $\beta_{{\bk}}$ or $E_{\vec{k}}$. 

Equations (\ref{SelfConsistEquMF}) allow to derive asymptotic relations. In the limit $\epsilon_{\bk}\to 0 $, we then have $E_{\bk}\to 0$ and $\beta_{\bk}\to \frac{c}{4T}$, ensuring the self-consistent solutions are well controlled in the $\epsilon_{\bk}\to 0$ regime that we are interested in. Near $\epsilon_{\bk}=0$ and for large $T \gg c$ ($\beta_{\vec{k}} \ll 1$), the renormalized spectrum is given by $E_{\bk} = \frac{\sqrt{1+\beta_{\bk}^2}}{1-\beta_{\bk}^2}\epsilon_{\bk} \approx \sqrt{1+3\beta_{\bk}^2}\epsilon_{\bk} \approx \sqrt{1+\frac{3c^2}{16T^2}}\epsilon_{\bk} $. The suppression of DOS is now given by 
\begin{equation*}
    \frac{\rho_F(\phi_0)}{\rho_F(\phi_0=0)} = \frac{1}{\sqrt{1+\alpha'^2}}, \quad \alpha'=\frac{3\sqrt{3}\phi_0\lambda^2 r_N}{2v_N^2T(r_dr_N-\phi_0^2)},
\end{equation*}
which is of the same form  as \equref{SuppressionAndAlpha}, found through the perturbative calculation presented in the main text and derived in \secref{sec:suppressiondos}.

When $T/c=1/4$, we have $\beta_{\bk}^2=1$ for $\epsilon_{\vec{k}}\rightarrow 0$, and \equref{SelfConsistEquMF} are not valid. At this point, the self consistent solutions open up a gap in $E_{\bk}$ when $\epsilon_{\bk}=0$. This gap follows by solving the equation $\beta_{\vec{k}}^2=1$. When $\epsilon_{\bk}=0$ and $\beta_{\bk}=1$, we also have $\tilde\epsilon_{\bk}=-\tilde\Delta_{\bk}$ [see \equref{SecondMFSelfConEq}] which gives $E_{\bk}=\sqrt{2}\tilde\epsilon_{\bk}$. For $T/c$ approaching $1/4$ from below, we find that $\beta_{\bk} \approx \frac{c}{4T}\left(1-\frac{1}{12}\frac{E_{\bk}^2}{T^2}\right)$. Thus the condition that $\beta_{\bk}^2=1$ gives us $E_{\bk} = \sqrt{12}T\sqrt{1-\frac{4T}{c}}$.

\begin{figure}[bt]
   \centering
    \includegraphics[width=\linewidth]{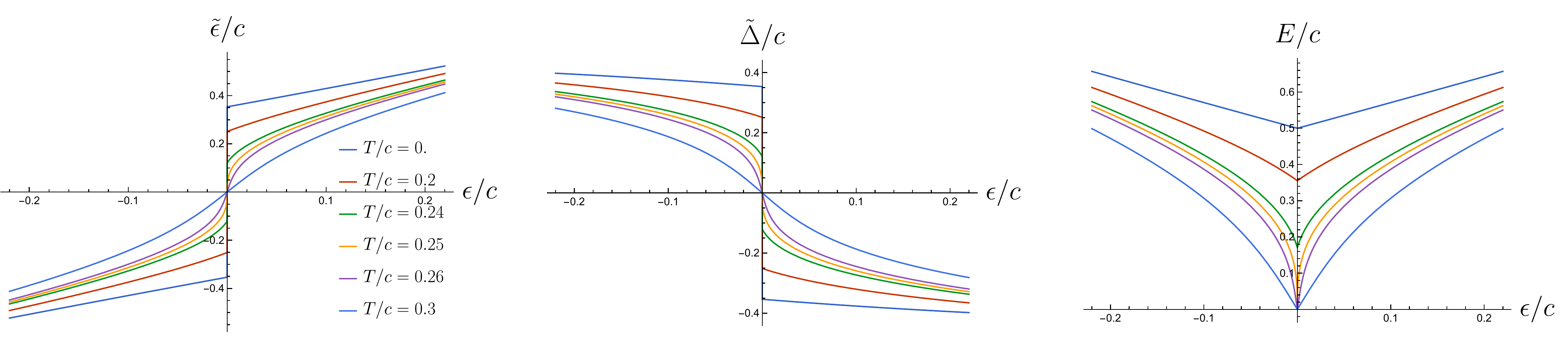}
    \caption{The self consistent solution for $\tilde\epsilon_{\bk},\tilde{\Delta}_{\bk}$ and $E_{\bk}$ as a function of $\epsilon_{\bk}$ for various temperatures. At $T/c=1/4$, the self-consistent solutions become non-analytic having an infinite slope at $\epsilon_{\bk}=0$, and a gap opens up as the temperature decreases. There is a discontinuity in $\tilde\epsilon_{\bk},\tilde\Delta_{\bk}$ at $\epsilon_{\bk}=0$, where the gap value has different signs for $\epsilon_{\bk}\rightarrow 0^-,0^+$.}
    \label{fig:selfconsistent}
\end{figure}

To summarize, for $T>c/4$, self consistent energy and gap ($\tilde\epsilon,\tilde\Delta$) are proportional to $\epsilon$. As $T$ approaches $c/4$ from above, the slope of proportionality approaches $\infty$ at $\epsilon=0$, and becomes non-analytic at $T=c/4$. Going below $T=c/4$, this non-analyticity at $\epsilon=0$ turns into a discontinuity at $\epsilon=0$, with the self consistent solutions developing a finite gap. The value of this gap at $T=0$ is given as $|\tilde\Delta|=|\tilde\epsilon| = \frac{|c|}{2\sqrt{2}}$. Figure \ref{fig:selfconsistent} illustrates the behavior obtained by numerical solution of the self-consistency equations.

\subsection{Zero energy-momentum transfer}
In this section, we consider the limit where the bosonic fields $\bN,\bd$ do not transfer any momentum or Matsubara frequency in the interaction ($q=0$ in $\mathcal{S}_c$). Additionally, we consider only the effect of $\mathcal{S}_2$ on the self energy to study the effect of the anomalous contribution. In this limit, we would like to analyze the self consistent solution of the Green's function up to all orders in $\lambda$ within the large-$N$ theory of the main text. The ansatz of the full Green's function is given by $G^{-1} = i\omega_n -\tilde{\epsilon}_k\gamma_z - \tilde{\Delta}_k\gamma_y$, since $\Sigma_3$ renormalizes only the anomalous term $\tilde\Delta_k$ and the spectrum $\tilde\epsilon_k$. 
We have 
\begin{align}
    G= \frac{i\omega_n+\tilde{\epsilon}_k\gamma_z+\tilde{\Delta}_k\gamma_y}{(i\omega_n)^2-\tilde{\epsilon}_k^2-\tilde{\Delta}_k^2}.
\end{align}
Thus the self-consistent analogue of $\Sigma_3$ in \equref{SelfEnergy3} becomes (where we have replaced the integration over $q$ by the $q=0$ value of the integrand, and $\tilde\phi_0=\phi_0\lambda^2r_N/v_N^2$)
\begin{align}
    \Sigma_3 = 6T\frac{\tilde\phi_0}{r_Nr_d-\phi_0^2}\frac{\tilde{\epsilon}_k\gamma_y+\tilde{\Delta}_k \gamma_z}{(i\omega_n)^2-\tilde{\epsilon}_k^2-\tilde{\Delta}_k^2}.
\end{align}
From the self-energy equation we get
\begin{align}
G^{-1}&=G_0^{-1}-\Sigma_3\\
    i\omega_n -\tilde{\epsilon}_{k}\gamma_z - \tilde{\Delta}_k\gamma_y &= i\omega_n - \epsilon_{\vec{k}} \gamma_z -6T\frac{\tilde\phi_0}{r_Nr_d-\phi_0^2}\frac{\tilde{\epsilon}_k\gamma_y+\tilde{\Delta}_k \gamma_z}{(i\omega_n)^2-\tilde{\epsilon}_k^2-\tilde{\Delta}_k^2}\\
    \tilde{\epsilon}_k &= \epsilon_{\bk} +Tc\frac{\tilde{\Delta}_k}{(i\omega_n)^2-\tilde{\epsilon}_k^2-\tilde{\Delta}_k^2}\label{eq:sceps}\\
     \tilde{\Delta}_k& = Tc\frac{\tilde{\epsilon}_k}{(i\omega_n)^2-\tilde{\epsilon}_k^2-\tilde{\Delta}_k^2},\label{eq:scdelta}
\end{align}
where $c=\frac{6\tilde\phi_0}{r_Nr_d-\phi_0^2}$. Right at the Fermi surface, $\epsilon_{\vec{k}}=0$, the self consistency equations reduce to 
\begin{align}
    \tilde{\epsilon}_k &= Tc\frac{\tilde{\Delta}_k}{(i\omega_n)^2-\tilde{\epsilon}_k^2-\tilde{\Delta}_k^2} \label{SCDiagrm1}\\
    \tilde{\Delta}_k& = Tc\frac{\tilde{\epsilon}_k}{(i\omega_n)^2-\tilde{\epsilon}_k^2-\tilde{\Delta}_k^2} \label{SCDiagrm2}\\
    \Longrightarrow  \tilde{\epsilon}_k &= T^2c^2\frac{\tilde{\epsilon}_k}{(\omega_n^2+\tilde{\epsilon}_k^2+\tilde{\Delta}_k^2)^2}
\end{align}
There are two possible solutions to \equsref{SCDiagrm1}{SCDiagrm2}. The first is $\tilde{\epsilon}_k=\tilde{\Delta}_k =0$; this is exactly what we find within perturbation theory. For a solution with $\tilde\epsilon_{k}\ne 0$ to exist, it must hold (assuming $\tilde{\epsilon}_k$, $\tilde{\Delta}_k \in \mathbb{R}$ as expected in the gauge that we use)
\begin{align}
    1=T^2\frac{c^2}{(\omega_n^2+\tilde\epsilon_{k}^2+\tilde\Delta_{k}^2)^2}<T^2\frac{c^2}{\pi^4T^4}
\end{align}
Thus, a non-zero solution only exists if $T<c/\pi^2\sim c/9$. As compared to Hartree-Fock, the critical temperature for a non-perturbative solution is lower.

\end{appendix}

\end{document}